\documentclass[a4paper,fleqn,usenatbib]{mnras}

\usepackage{newtxtext}

\usepackage[T1]{fontenc}
\usepackage{ae,aecompl}

\usepackage{graphicx}	
\usepackage{amsmath}	
\usepackage{amssymb}	

\def\hi{H\,{\sc i}}

\def\kms{km~s$^{-1}$}
\def\msun{$M_{\odot}$}

\long\def\symbolfootnote[#1]#2{\begingroup%
\def\thefootnote{\fnsymbol{footnote}}\footnote[#1]{#2}\endgroup} 

\title[Synthetic data products for \hi\ galaxy surveys]{Synthetic data products for future \hi\ galaxy surveys: a tool for characterising source confusion in spectral line stacking experiments}

\author[E. C. Elson et al.]{
E. C. Elson,$^{1,2}$\thanks{E-mail: elson.e.c@gmail.com (ECE)}
S. L. Blyth,$^{1,2}$
A. J. Baker,$^{3}$
\\
$^{1}$Astrophysics, Cosmology and Gravity Centre (ACGC), Department of Astronomy, University of Cape Town, Private Bag X3,
Rondebosch 7701, South Africa.\\
$^{2}$Square Kilometre Array South Africa, 3rd Floor, The Park, Park Road, Pinelands, 7405, South Africa.\\
$^{3}$Department of Physics and Astronomy, Rutgers, The State University of New Jersey, 136 Frelinghuysen Road, Piscataway, NJ 08854-8019, USA.}

\pubyear{2016}

\begin{document}
\label{firstpage}
\pagerange{\pageref{firstpage}--\pageref{lastpage}}
\maketitle

\begin{abstract}
Much of our current understanding of neutral, atomic gas in galaxies comes from radio observations of the nearby Universe.  Until the next generation of instruments allow us to push to much higher redshifts, we must rely mostly upon theoretical models of galaxy formation to provide us with key insights into the likely cosmic evolution of \hi\ in the Universe, and its links to molecular clouds and star formation.  In this work, we present a new set of methods to convert mock galaxy catalogues into synthetic data cubes containing model galaxies with realistic spatial and spectral \hi\ distributions over large cosmological volumes.  Such synthetic data products can be used to guide observing and data handling/analysis strategies for forthcoming \hi\ galaxy surveys.  As a demonstration of the potential use of our simulated products we use them to conduct several mock \hi\ stacking experiments for both low and high-redshift galaxy samples.  The stacked spectra can be accurately decomposed into contributions from target and non-target galaxies, revealing in all co-added spectra  large fractions of contaminant mass due to source confusion.  Our results are consistent with similar estimates extrapolated from $z=0$ observational data.  The amount of confused mass in a stacked spectrum grows almost linearly with the size of the observational beam, suggesting potential over-estimates of $\Omega_{\mathrm{HI}}$ by some recent \hi\ stacking experiments.  Our simulations will allow the study of subtle redshift-dependent effects in future stacking analyses.
\end{abstract}


\section{Introduction}
Observations of neutral, atomic hydrogen (\hi) in emission have always  been limited to the nearby Universe.  Owing to the intrinsic faintness of the \hi\ emission line, galaxies beyond a few hundred megaparsecs become difficult to image directly in a reasonable amount of time.  This in many ways prohibits a quantitative study of galaxy evolution  given that total cold gas content is one of the key drivers of star formation.  Most of our current understanding of \hi\ in galaxies comes from nearby galaxy surveys such as the \hi\ Parkes All-Sky Survey (HIPASS, \citealt{HIPASS}), the \hi\ Jodrell All Sky Survey (HIJASS, \citealt{HIJASS}), The \hi\ Nearby Galaxy Survey (THINGS, \citealt{THINGS_walter}) and the Arecibo Fast Legacy ALFA Survey (ALFALFA, \citealt{ALFALFA}).  \hi\ mass functions (HIMFs) measured from these surveys are used to evaluate the total mass density of neutral hydrogen, $\Omega_{\mathrm{HI}}$, of the local Universe (e.g.~\citealt{zwaan_HIMF,Martin_2010}).

Until we have observational constraints on the cosmic evolution of \hi, we must rely upon theory to provide us with insights into the  evolving properties of cold gas, in particular $\Omega_{\mathrm{HI}}$.  In recent years, semi-analytic models of galaxy formation have focused on methods of predicting the cold gas properties of galaxies.  \citet{obresch_2009a}  present a simulation of the cosmic evolution of the atomic and molecular phases of the cold hydrogen gas in $\sim~3\times10^7$ galaxies.  They provide results for the \hi\ and $\mathrm{H_2}$ mass functions, the CO luminosity function, the cold gas mass-diameter relation, and the Tully-Fisher relation; all of which match observational data from the nearby Universe.  They also present high-redshift predictions for cold gas disc sizes and the Tully-Fisher relation, both of which appear to change significantly with lookback time.  \citet{Lagos_2011a} use self-consistent models to predict the \hi\ and $\mathrm{H_2}$ content of galaxies,  successfully matching nearby Universe observations, as well as high-redshift observations.  They predict the HIMF to evolve weakly with redshift, with the number density of massive galaxies decreasing with increasing redshift.  The updated models from \citet{Lagos_2014a} predict a modest evolution of the cosmic \hi\  density for $z<3$, with the \hi\ density being dominated by galaxies with stellar mass $M_*<10^9$~\msun.  \citet{Popping_2014} use semi-analytic models with pressure-based and metallicity-based scenarios for the formation of molecules.  Both recipes predict that galaxy gas fractions remain high from $z\sim 3-6$ and drop rapidly at lower redshift.

These models all emphasise the crucial role played by \hi\ in galaxy formation and evolution.  In order to fully utilise their predictive power, they need to be converted into a format that can be treated and handled in the same ways as real data.  To this end, we have developed a set of tools that use an input catalogue of evaluated galaxy properties to generate synthetic data cubes for large cosmological volumes.   The data cubes contain detailed models of the spectral and spatial distributions of \hi\ in galaxies.  Such synthetic products can be used to guide the planning of radio and supporting multi-wavelength  observations for future \hi\ galaxy surveys, as well as guide development of calibration, imaging and analysis methods.  In this work, we present the techniques used to construct our synthetic data products.  We also present an application of them in the form of mock \hi\ stacking experiments for both low and high-redshift galaxy samples. 

The next generation of large \hi\ galaxy surveys aims to track the cosmic evolution of \hi\ over large areas and/or out to $z>1$.  However, much like current-generation surveys at intermediate redshifts, all of them will struggle to directly image high-redshift galaxies in a reasonable amount of time.   \hi\  spectral line stacking is the preferred method for measuring  the total \hi\ content of populations  of galaxies that are too faint to detect individually.    The process uses  known spatial locations and redshifts of a sample of galaxies to extract their \hi\ spectra from a data cube.  The individual \hi\ spectra are aligned and then co-added to yield a single stacked (or co-added) spectrum that is representative of the \hi\ line emission from the entire galaxy sample. 

An unappreciated shortcoming of the \hi\ stacking method is source confusion (e.g.,~\citealt{Jones_2015a,Jones_2015b}). Depending on the level of source crowding in a data cube, an extracted spectrum can easily be contaminated by flux from other nearby galaxies.  Any co-added spectrum based on contaminated spectra will overestimate the true total \hi\ mass of a galaxy sample.  For their Parkes observations of galaxies spanning the redshift range $0.04 < z < 0.13$, for example, \citet{delhaize_2013} estimate any one galaxy to be confused, on average, with seven others in the 2dFGRS optical sample.  This is primarily due to the large angular size of the Parkes beam, 15.5~arcmin.  More recently  \citet{Jones_2015a,Jones_2015b} used the ALFALFA correlation function to develop an analytic model to predict the amount of confused flux in a stacked \hi\ spectrum.  For a Parkes stacking experiment; their model predicts the average amount of confused \hi\ mass per galaxy to be $\sim~1.3\times10^{10}~h_{70}^{-2}$~\msun\ for a beam size of 15.5~arcmin.

Our synthetic \hi\ data cubes are ideally suited to carrying out mock experiments that allow us to accurately and reliably quantify the  inherent uncertainties of the \hi\ stacking method.  We use the results from the stacking experiments presented in this work to quantify the rates of source confusion in a series of low-redshift \hi\ stacking experiments, as well as several high-redshift stacking experiments that may be carried out with the Square Kilometre Array (SKA) and its precursors.  

The layout of this paper is as follows: In Section~\ref{sec:simulations}, we present the details of converting mock galaxy catalogues into synthetic \hi\ data cubes.  We show in Section~\ref{sec:flux_decomp} how our simulations can be used to precisely decompose the flux in a spectrum extracted from a synthetic cube into contributions from a target galaxy of interest, as well as contributions from other non-target galaxies.  In Section~\ref{sec:mock_stacks}, we carry out and present the results of a suite of low-and high-redshift \hi\ stacking experiments.  For each experiment we study the breakdown of the co-added \hi\ mass into contributions from target and non-target galaxies.  We discuss our results and the insights gained from our mock experiments in Section~\ref{sec:discussion}, and in Section~\ref{sec:conclusions} we present our conclusions.  Throughout this work we have assumed a $\Lambda$CDM cosmology with a Hubble constant $H_{\mathrm{0}}=67.3$~\kms~Mpc$^{-1}$, $\Omega_{\Lambda}=0.685$ and $\Omega_{\mathrm{M}}=0.315$ \citep{Planck_cosmology}.

\section{Simulations}
\label{sec:simulations} 
The  simulated data products presented  and used in this work are based on the catalogue of evaluated galaxy properties from \citet{obresch_2014}.  The catalogue spans a sky area of 10-by-10 degrees and the redshift range $z=0-1.2$.  For several millions of galaxies, it presents detailed \hi\ properties as well as auxiliary optical properties.  For 21~cm peak flux densities above $1\mu$Jy, it is complete down to an \hi\ mass of 10$^8$~\msun.  The catalogue is based on the SKA Simulated Skies semi-analytic simulations (S$^3$-SAX), delivered in 2009 as part of the European SKA Design Studies (SKADS), and therefore on the physical models described in \citet{obresch_2009a,obresch_2009b,obresch_2009c}.  These models are able to assign realistic masses and sizes to \hi\ discs, and evaluate the characteristic properties of their \hi\ emission lines.  

Our simulation method involves realistically modelling, in a fully three-dimensional manner, the spatial and spectral distribution of the \hi\ line emission for any galaxy from the \citet{obresch_2014} catalogue.  In the sections that follow, we describe how we create these models, as well as how they are brought together to produce bespoke artificial \hi\ data cubes representing large cosmological volumes. 

\subsection{Galaxy models}
A rotation curve (circular velocity profile) and a parameterisation of the radial distribution of \hi\ mass for a galaxy are the main requirements for generating a full three-dimensional model - i.e., a mini  data cube - of the spatial and spectral distribution of its \hi\ line emission.

\subsubsection{HI mass profiles}
Our adopted parameterisation for the azimuthally-averaged radial distribution of HI mass is:
\begin{equation}
\Sigma_{\mathrm{HI}}(R) = {A\exp(-R^2/2h^2) \over 1 + \beta\exp(-1.6R^2/2h^2)}.
\label{eq:HIprofile}
\end{equation}

$A$ is a normalisation factor used to control the total \hi\ mass.  The numerator is a two-dimensional Gaussian with standard deviation $h$.  The $\beta$ parameter in the denominator controls the central concentration of \hi.  Values of $\beta>0$ yield central \hi\ depressions.  Equation~(\ref{eq:HIprofile}) is capable of matching \hi\ distributions observed in nearby galaxies (e.g.,~\citealt{bigiel_mass_distributions, martinnson_thesis}). 

\begin{figure}
	\includegraphics[width=\columnwidth]{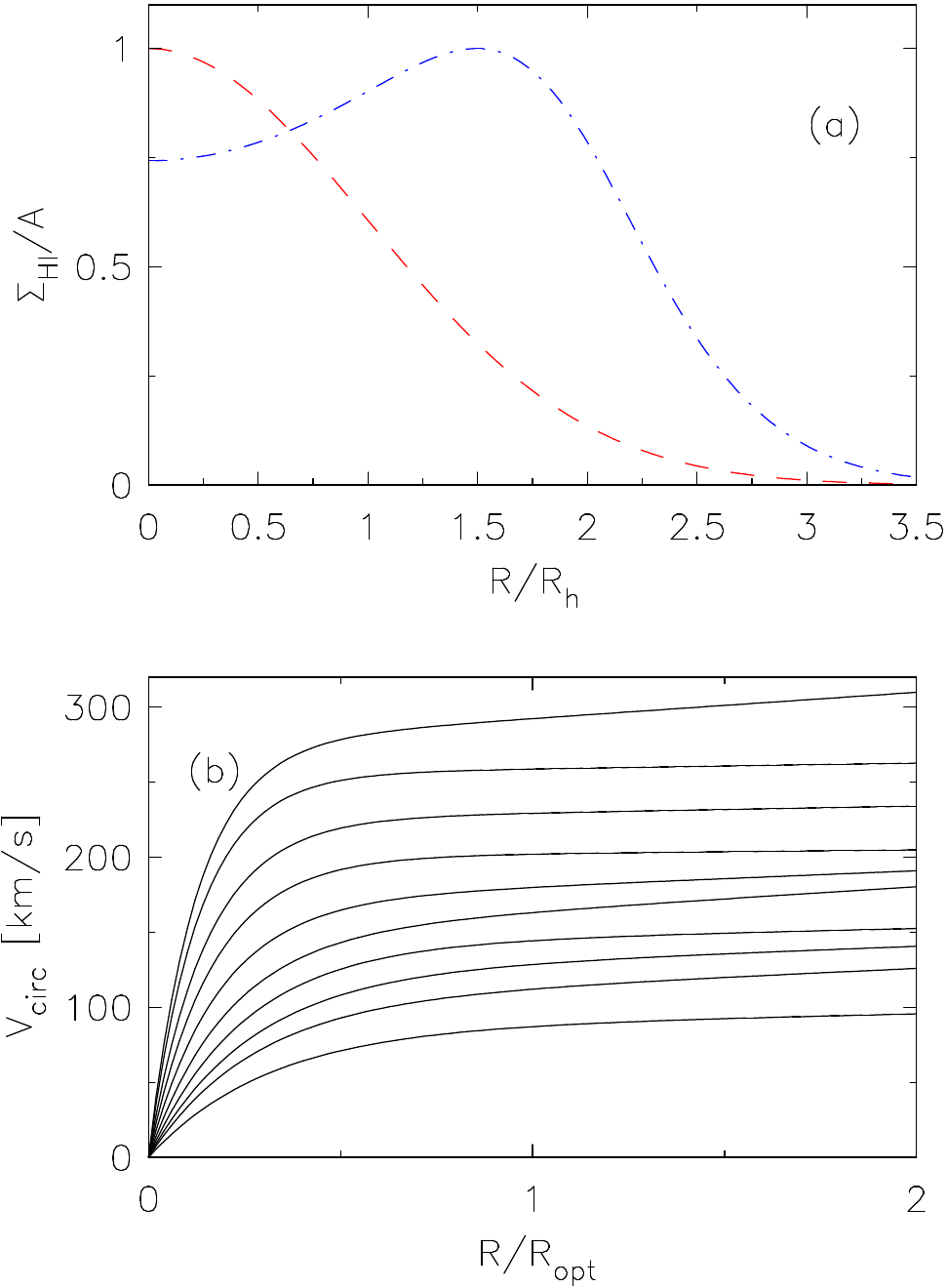}
    \caption{Panel (a): \hi\ mass profiles for parameter pairs $(h, \beta)~=~(R_h, 0.0),~(R_h, 10.0)$ shown as red dashed and blue dot-dashed curves, respectively.  Both mass profiles have been scaled to have a maximum value of unity, for the sake of clarity.  $R$ is shown in units of $R_h$.  Panel~(b):  Polyex model fits to template rotation curves parameterised as functions of optical radius.  The mean $I$-band absolute magnitudes of the various models (from top to bottom) are given in the first column of Table~\ref{tab:polyex_parameters}.  Each of the 3D galaxy models we generate in this work is based on one of  these model rotation curves.}
    \label{fig:example_profiles}
\end{figure}

To construct an \hi\ mass profile for a galaxy in the \citet{obresch_2014} catalogue we convert its evaluated apparent \hi\ half-mass radius along the major axis, $R_{\mathrm{HI}}^{\mathrm{half}}$, into an exponential disc scale length, $R_h$.  We then set $h=R_h$ in equation~(\ref{eq:HIprofile}).  $\beta$ is set equal to $R_{\mathrm{mol}}^{\mathrm{c}}$ from the \citet{obresch_2014} catalogue, which (for that paper's similar but not identical mass profile parameterisation) is the extrapolated central $\mathrm{H_2}$/\hi\ mass ratio for an exponential disc.  The first panel in Fig.~\ref{fig:example_profiles} shows the \hi\ mass profiles for two galaxies from the catalogue.  In this work, all galaxies are modelled out to a radius $R=3.5h$.  

\subsubsection{Circular velocity profiles}
Circular velocity profiles of galaxies (rotation curves) are modelled using the template rotation curves for disc galaxies presented in \citet{catinella_2006}, who use a homogeneous sample of $\sim 2200$ low-redshift disc galaxies to construct average rotation curves in separate $I$-band luminosity classes.   Their template rotation curves are fitted with the analytic function:
\begin{equation}
V_{\mathrm{PE}}(R)=V_{0}\left(1-e^{-R/R_{\mathrm{PE}}}\right)\left( 1+{\alpha R\over R_{\mathrm{PE}}}\right).
\label{eq:vrot}
\end{equation}
This is the so-called Polyex model from \citet{giovanelli_haynes_2002}.  $V_0$, $R_{\mathrm{PE}}$ and $\alpha$, respectively, determine the amplitude, the exponential scale of the inner rotation curve, and the slope of the outer rotation curve.  \citet{catinella_2006} point out that the Polyex model is an empirical expression that nicely fits a large variety of rotation curve shapes, including those declining at large radii.   

For various luminosity classes, Table~2 from \citet{catinella_2006} presents Polyex model fits to template rotation curves parameterised as functions of optical radius.  It  is this information that we use  to construct rotation curves for galaxies based on their evaluated $R$-band absolute magnitudes.  We convert from $R$- to $I$-band absolute magnitudes using $M_{\mathrm{I}}=M_{\mathrm{R}}-0.37$, based on the evaluated optical magnitudes for $z<0.1$ galaxies with stellar mass M$_*>10^{10}$~\msun\ in the \citet{Duffy_2012} semi-analytic simulations.  For the sake of convenience, a subset of the information in Table~2 from \citet{catinella_2006} is reproduced here in Table~\ref{tab:polyex_parameters}.  Plots of the various Polyex model fits are shown in the second panel of Fig.~\ref{fig:example_profiles}.

\begin{table}
	\centering
	\caption{Polyex model fits to template rotation curves as functions of optical radius for ten $I$-band defined luminosity classes, taken from Table~2 of \citet{catinella_2006}.  Column 1 gives average $I$-band absolute magnitude.  Columns 2 - 4 give, respectively, the amplitude (in \kms) of the outer rotation curve, the exponential scale of the inner rotation curve (in units of optical radius), and the slope of the outer rotation curve. }
	\label{tab:polyex_parameters}
	\begin{tabular}{cccc} 
		\hline
		$\left<M_I\right>$	&	$V_0$		&	$R_{\mathrm{PE}}/R_{\mathrm{opt}}$	&	$\alpha$			\\
		\hline	
		-23.76				&	$275\pm6$	&	$0.126\pm0.007$					&	$0.008\pm0.003$		\\
		-23.37				&	$255\pm2$	&	$0.132\pm0.003$					&	$0.002\pm0.001$		\\
		-22.98				&	$225\pm1$	&	$0.149\pm0.003$					&	$0.003\pm0.001$		\\
		-22.60				&	$200\pm1$	&	$0.164\pm0.002$					&	$0.002\pm0.001$		\\
		-22.19				&	$170\pm1$	&	$0.178\pm0.003$					&	$0.011\pm0.001$		\\
		-21.80				&	$148\pm2$	&	$0.201\pm0.004$					&	$0.022\pm0.002$		\\
		-21.41				&	$141\pm2$	&	$0.244\pm0.005$					&	$0.010\pm0.003$		\\
		-21.02				&	$122\pm2$	&	$0.261\pm0.008$					&	$0.020\pm0.005$		\\
		-20.48				&	$103\pm2$	&	$0.260\pm0.008$					&	$0.029\pm0.005$		\\
		-19.38				&	$85\pm5$	&	$0.301\pm0.022$					&	$0.019\pm0.015$		\\
		\hline
	\end{tabular}
\end{table}

\subsubsection{Three-dimensional modelling}
To generate \hi\ data cubes based on the radial profile parameterisations presented in equations~(\ref{eq:HIprofile}) and (\ref{eq:vrot}), we use our own routine that assumes axisymmetry and models a galaxy as a collection of many \hi\ clouds. For each galaxy, the inclination and position angle of the \hi\ disc must be provided together with a rotation curve. Additional inputs required for the model are the \hi\ velocity dispersion and the \hi\ mass profile as functions of radius. Assuming an infinitely thin \hi\ disc, the routine uses the \hi\ mass profile and rotation curve as probability density functions to generate many random numbers representing the spatial and spectral coordinates of individual \hi\ clouds.  A small random velocity is added to the galactocentric rotation velocity of each cloud.  These random velocities are extracted from a Gaussian centred at 0~\kms\ and with a standard deviation equal to the specified \hi\ velocity dispersion at the galactocentric radius at which the cloud is placed.  Finally, the routine uses the position and velocity of the cloud to place it at the appropriate location within the data cube.

\begin{figure*}
	\includegraphics[width=2\columnwidth]{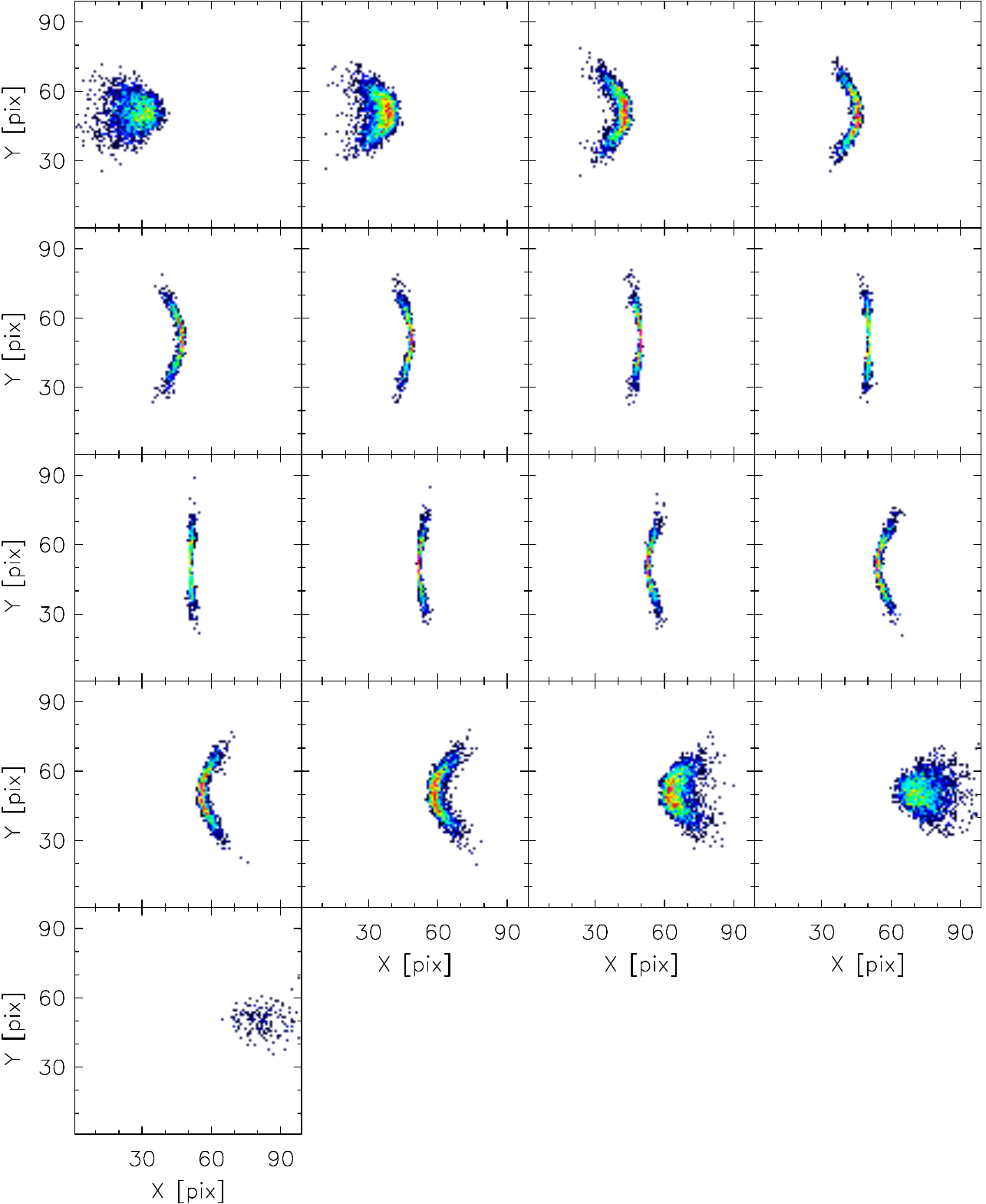}
    \caption{Channel maps from an \hi\ data cube model for a single galaxy in the \citet{obresch_2014} catalogue. The model has its rotation curve represented by the bottom curve in Fig.~\ref{fig:example_profiles}(b).   Channels in the cube are  5~\kms\ wide; every second channel is shown here.  This model has not been flux-calibrated; colours in the maps represent only the relative distribution of \hi\ flux in a particular channel.}
    \label{fig:eg_model_channel_maps}
\end{figure*}

\begin{figure*}
	\includegraphics[width=1.5\columnwidth]{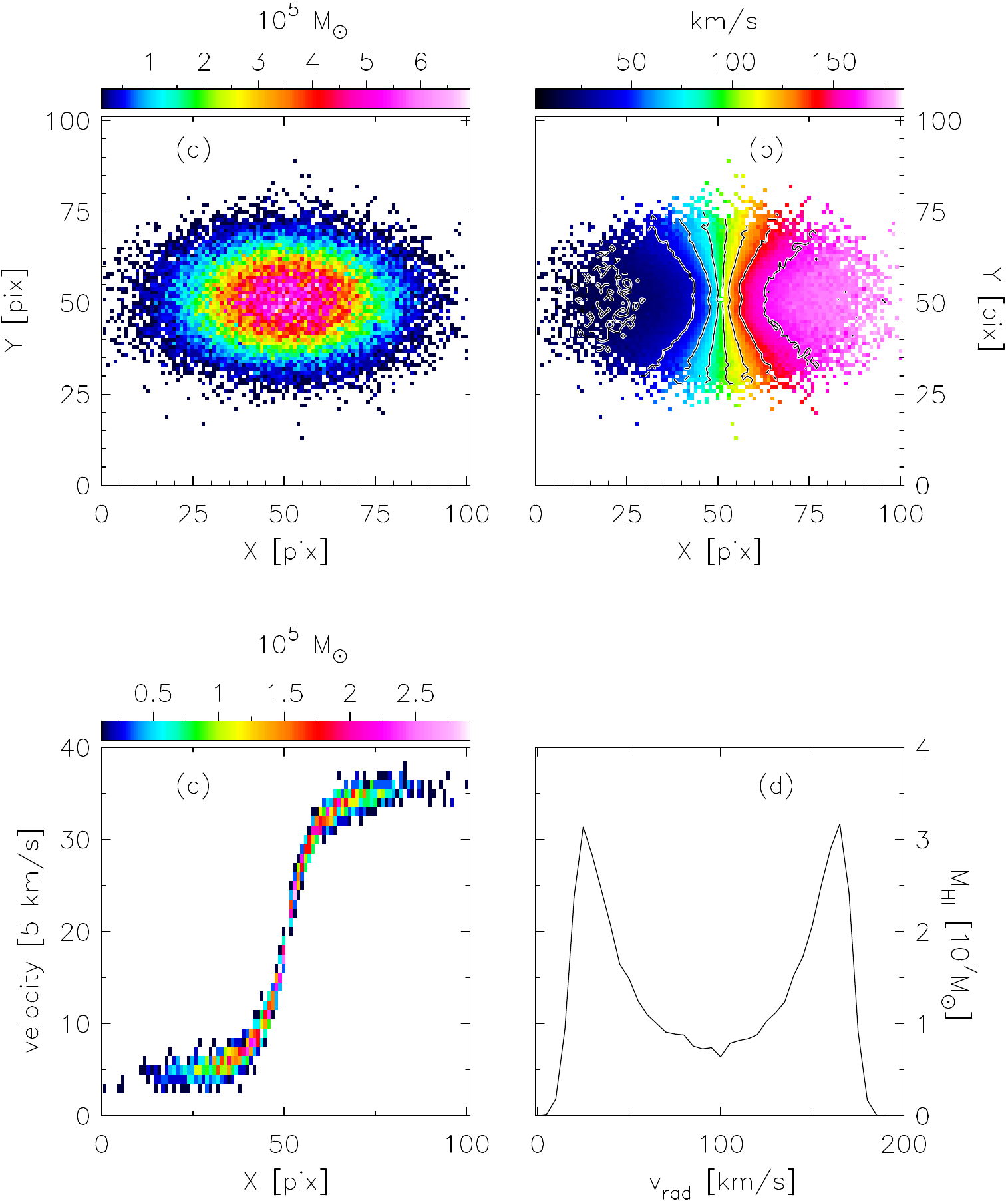}
    \caption{\hi\ data products generated from the model cube shown in Fig.~\ref{fig:eg_model_channel_maps}.  Panel~(a): total intensity map showing the spatial distribution of \hi\ mass.  Panel~(b): intensity-weighted mean velocity field.  Iso-velocity contours are spaced by 20~\kms.  Panel~(c): position-velocity slice extracted along the major axis of the galaxy.  Panel~(d): global profile showing the total \hi\ mass in each channel.  The galaxy has an \hi\ mass $M_{\mathrm{HI}}=5.2\times 10^{8}$~\msun\ and inclination $i=55$~degrees.  Its rotation curve is represented by the bottom curve in Fig.~\ref{fig:example_profiles}(b).}
    \label{fig:eg_model_data_products}
\end{figure*}

Figure~\ref{fig:eg_model_channel_maps} shows  channel maps for a single galaxy model.  This galaxy has an \hi\ mass $M_{\mathrm{HI}}=5.2\times 10^{8}$~\msun, with  $\beta=0.8$ - resulting in an \hi\ mass profile very similar to the red-dashed curve in Fig.~\ref{eq:HIprofile}(a).  The $I$-band absolute magnitude of the galaxy is $M_I=-19.2$; its rotation curve is represented by the bottom curve in Fig.~\ref{fig:example_profiles}.  The \hi\ disc is inclined at 55~degrees.  All galaxy models are created with a position angle of $270$~degrees\symbolfootnote[5]{Measured anti-clockwise from north to the receding major axis.} and spatial dimensions of 101-by-101 pixels.  A variable pixel scale is used to ensure that each galaxies has its \hi\ major axis spanning  a full 101 pixels in the cube (e.g.,~Fig.~\ref{fig:eg_model_data_products}a).  A common channel width of 5~\kms\ is used for all models.  Each model has a number of channels equal to $2V_{\mathrm{max}}\sin (i)/5$, rounded up to the nearest integer, where $V_{\mathrm{max}}$ is the maximum circular rotation speed (in \kms) given by the rotation curve of the galaxy.  For completeness, Fig.~\ref{fig:eg_model_data_products} shows the following data products generated from the Fig.~\ref{fig:eg_model_channel_maps} data cube: a) total intensity map, b) intensity-weighted mean velocity field, c) major axis position-velocity slice, and d) global profile (\hi\ spectrum).

\subsection{Lightcone models}
The galaxy models described thus far serve as the building blocks for our simulations.  For a user-specified subset of a lightcone from the \citet{obresch_2014} catalogue, we model each galaxy in the ways described above.  All of the models are then combined to form the full-size synthetic cube.  

\begin{figure*}
	\includegraphics[width=2\columnwidth]{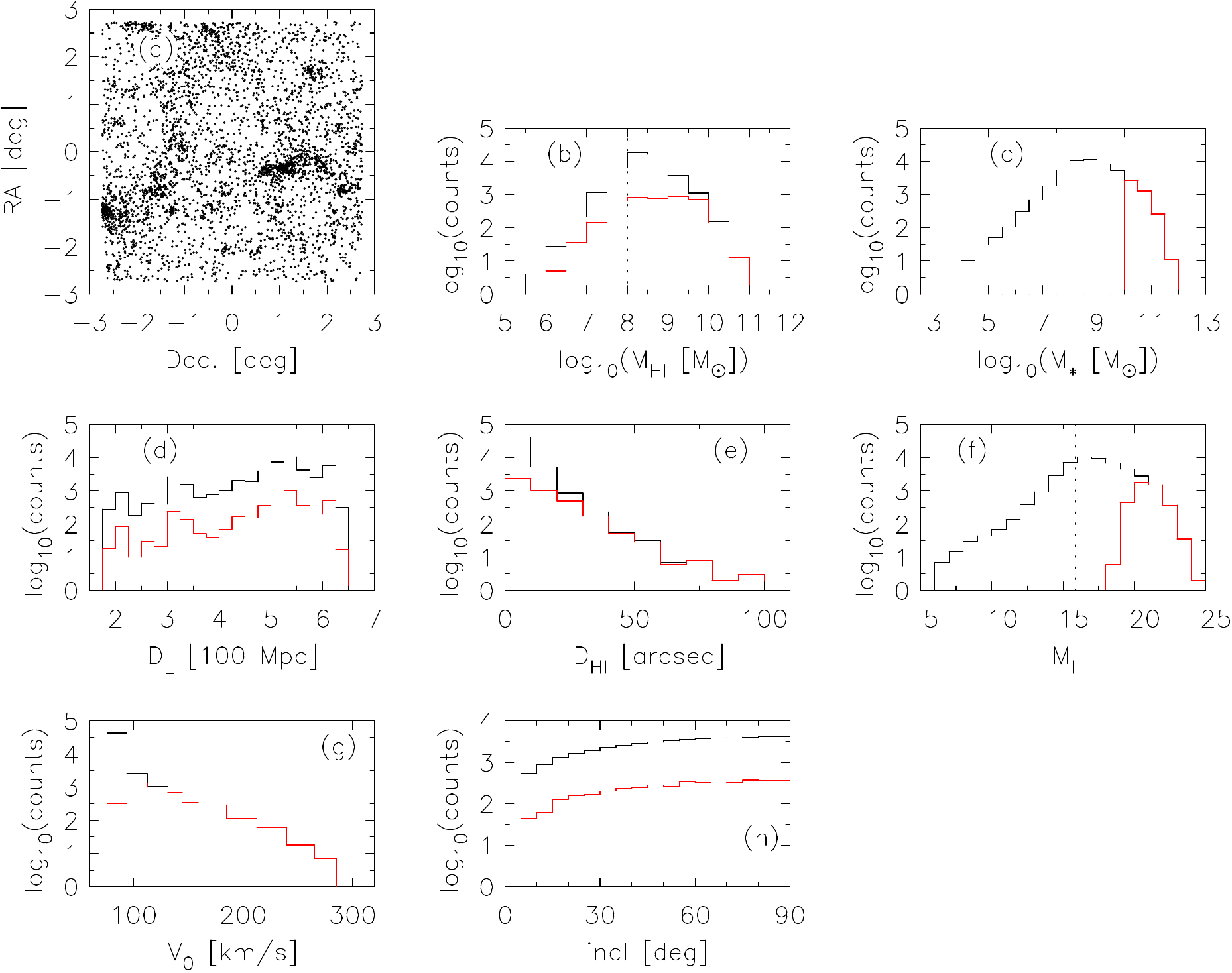}
    \caption{Evaluated galaxy properties for a 30 square degree subset of the \citet{obresch_2014} catalogue.  This subset, spanning the redshift range $z=0.04 - 0.13$, contains 48~234 galaxies and was used to generate a synthetic data cube with channel maps shown in Fig.~(\ref{fig:lightcone_channel_maps}).  The black histograms correspond to the full sample of 48~234 galaxies, whereas the red histograms are for the subset of 3583 galaxies with stellar mass $M_*>10^{10}$~\msun\ that are used to produce the co-added spectra in Fig.~\ref{fig:lowz_stacks}.  Panel~(a): (RA, Dec.) positions of a random subset of 4000 of the 48~234 galaxies.  Panel~(b): distribution of \hi\ masses.  Total \hi\ mass is  $2.81\times 10^{13}$~\msun\ for the full sample.  Panel~(c): distribution of stellar masses.  Panel~(d): distribution of luminosity distances.  Panel~(e): distribution of \hi\ diameters.  Panel~(f): distribution of $I$-band absolute magnitudes.  Panel (g): distribution of $V_{\mathrm{0}}$ parameters (see eqn~\ref{eq:vrot}), indicative of maximum rotation speed.  Panel~(h): distribution of galaxy inclinations.  The vertical dotted lines in panels b, c, f indicate the masses above which the simulation is complete.}
    \label{fig:lightcone_panel_plot}
\end{figure*}

As an example of this process, consider a particular catalogue subset spanning $\sim~30$ square degrees on the sky and the redshift range $z=0.04 - 0.13$\symbolfootnote[3]{Corresponding to the luminosity distance range 183~Mpc - 633~Mpc.}.  This volume contains 48~234 galaxies with a total \hi\ mass of $2.81\times 10^{13}$~\msun.  Some of the  properties of this galaxy sample are shown in Fig.~\ref{fig:lightcone_panel_plot}, including all of the information required to generate a three dimensional model of each galaxy.

Each model cube, which is created to have 101 by 101 pixels in right ascension and declination, is appropriately re-gridded before being placed into the full-size synthetic cube at the (RA, Dec) position specified by the catalogue.  The \citet{obresch_2014} catalogue provides an evaluation of the apparent \hi\ radius along the major axis of a galaxy out to an \hi\ mass surface density of 1~\msun~pc$^{-2}$.  This quantity together with a specified pixel scale for the full-size cube determines the number of spatial pixels the re-gridded model will span.  Each model also has its major axis position angle randomised before being placed into the full-size cube.  

The galaxy models are created using a velocity axis with a fixed channel of width $dv=5$~\kms, whereas the final full-size cube has a frequency axis of fixed channel width, $df$, specified by the user.  The velocity interval, $dv$, corresponding to a frequency interval, $df$, increases with redshift as follows:

\begin{equation}
dv={df\times c(1+z)\over f_{\mathrm{emit}}},
\label{eq:df_dv}
\end{equation}
where $c$ is the speed of light in a vacuum, $z$ is the galaxy redshift and $f_{\mathrm{emit}}$ is the rest frequency of \hi.  In practice this means that a particular galaxy will span fewer channels at a high redshift than it would at a low redshift.  To account for this the velocity axis of each model cube is re-gridded according to equation~(\ref{eq:df_dv}) in order to ensure that it spans the correct number of channels at its specified redshift. 

Having created a full-size cube containing the \hi\ line emission of the galaxies, it is convolved on a channel-by-channel basis with a user-specified point-spread-function (PSF).  In this work we use Gaussian PSFs.  Figure~\ref{fig:lightcone_channel_maps} shows the channel maps of the full-size cube for the $\sim~30$ square degree subset of the \citet{obresch_2014} catalogue mentioned above. The cube has a spatial resolution of 15~arcmin, a pixel scale of 30~arcsec and a channel width of 62.5~kHz ($\sim 14.3$~\kms at the mean redshift of $z=0.095$).  This version of the cube contains no noise, only spatially-smoothed \hi\ line emission.  An \hi\ total intensity map generated from this cube is shown in Fig.~{\ref{fig:lightcone_mom0}.  

In terms of real data, our cubes represent continuum-subtracted CLEANed data cubes (as opposed to dirty  cubes) that have been restored with a Gaussian approximation of the PSF main lobe.  Furthermore, our cubes are primary-beam-corrected, by design.  There is no need for us to apply any sort of weighting to  spectra extracted from our cubes. 

\begin{figure*}
	\includegraphics[width=2\columnwidth]{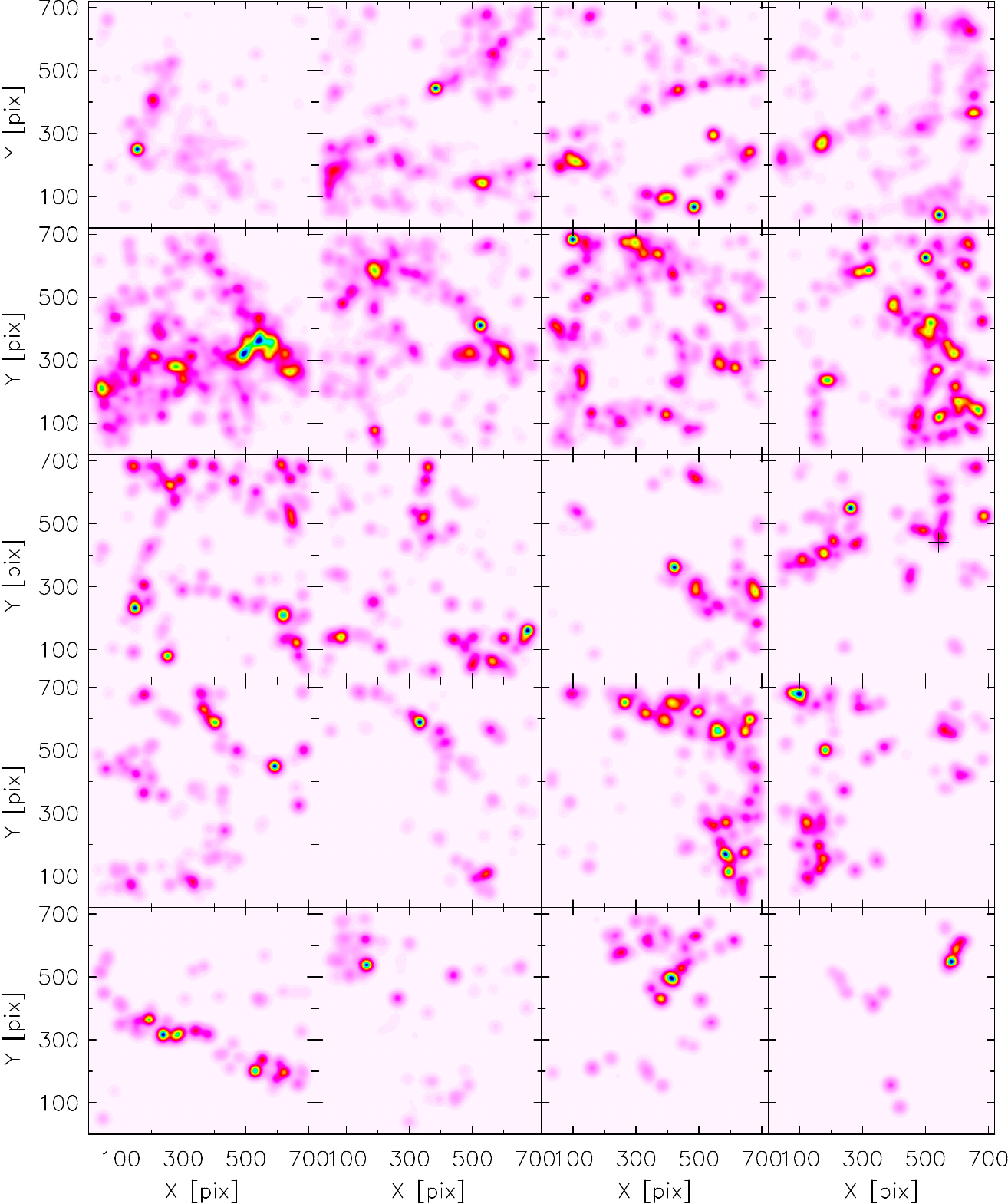}
    \caption{Channel maps showing the \hi\ line emission (no noise) of a cube spanning $\sim~30$ square degrees on the sky and a redshift range $z=0.04-0.13$.  The cube contains 48~234 galaxies and has a spatial resolution of 15~arcmin, a pixel scale of 30~arcsec and a channel width of 62.5~kHz ($\sim 14.3$~\kms at the mean redshift of $z=0.085$).  Each panel shown here represents the sum of 10 adjacent channels in the cube, spanning a velocity range  $\sim 143$~\kms.  Panels are spaced by $\sim 573$~\kms.  The total intensity map created from this cube is shown in Fig.~\ref{fig:lightcone_mom0}.  The black cross shown in the right-most panel of the middle row marks the centre of the sub-cube extracted and used to produce the spectra shown in Fig.~\ref{fig:15_arcmin_spectra}, as well as the spectra shown in Fig.~\ref{fig:spectra_at_various_res}.  Most of the emission in the sub-cube is contained within this channel map.}
    \label{fig:lightcone_channel_maps}
\end{figure*}

\begin{figure}
	\includegraphics[width=1\columnwidth]{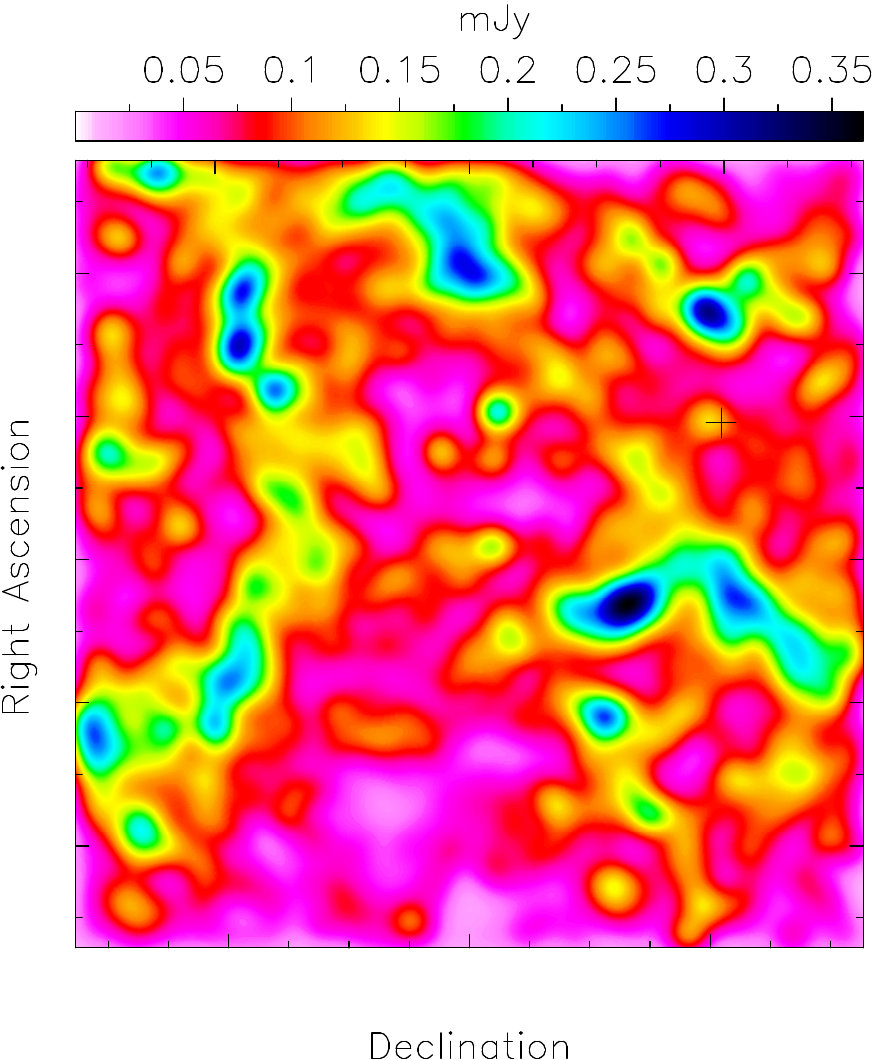}
    \caption{\hi\ total intensity map for the full-size synthetic cube with channel maps shown in Fig.~\ref{fig:lightcone_channel_maps}.  The map spans $\sim~30$ square degrees and contains the \hi\ line emission from 48~234 galaxies.  All of the emission in the map amounts to $44.33$~Jy.  The distribution of flux in this map should be compared to the true (RA, Dec) distribution of the galaxies shown in Fig.~\ref{fig:lightcone_panel_plot}a.  The black cross symbol marks the centre of the sub-cube extracted and used to produce the spectra shown in Fig.~\ref{fig:15_arcmin_spectra}, as well as the spectra shown in Fig.~\ref{fig:spectra_at_various_res}.} 
    \label{fig:lightcone_mom0}
\end{figure}

 \section{Quantifying flux confusion}
\label{sec:flux_decomp}
A major concern for \hi\ spectral line stacking experiments is the degree to which the spectrum of a particular target galaxy is potentially contaminated by emission from other nearby galaxies.  A very powerful application of our simulated cubes is to calculate with a high degree of accuracy and reliability the amount of contaminant emission present in any extracted spectrum.  In this section we provide a demonstration of how the flux in a particular sub-volume extracted from a large cube can be decomposed into contributions from the target galaxy as well as contaminant emission from other galaxies.  

For this demonstration we use the synthetic cube presented in Fig.~\ref{fig:lightcone_channel_maps} to extract a sub-cube centred on a particular galaxy with evaluated  \hi\ mass $1.51\times 10^{9}$~\msun.  The spatial position of this galaxy is marked with a cross in Figs~\ref{fig:lightcone_channel_maps}, \ref{fig:lightcone_mom0}.   At a redshift of 0.082, most of the flux from the galaxy is contained in the channel map shown in the right-most panel in the middle row of Fig.~\ref{fig:lightcone_channel_maps}.  Centred on the position of this galaxy we extract from the full-size synthetic cube a sub-cube spanning an spatial area of 15~arcmin$\times$~15~arcmin (30 pixels~$\times$~30 pixels) and a velocity range of 600~\kms.  

Summing the flux in each channel of the sub-cube yields the spectrum shown as the thick grey curve in panel (a) of Fig.~\ref{fig:15_arcmin_spectra}.  The spatial distribution of this emission is shown as the \hi\ total intensity map in panel (c).  The emission in the sub-cube is made up of target and non-target galaxy emission.  The flux from the target galaxy alone is represented by the green spectrum in panel (a) and the total intensity map in panel (d).  An \hi\  mass of $9.11\times 10^8$~\msun\ is associated with the target galaxy, which is only a small fraction of the total amount of  \hi\ mass in the sub-cube, $7.34\times 10^9$~\msun.  The difference is due to contaminant emission from other nearby galaxies.  The total intensity map in panel (a) is generated from a version of the sub-cube that has \emph{not} been spatially smoothed; it  therefore  represents the true brightness distribution of the galaxies in the sub-cube.  The target galaxy with \hi\ mass $9.11\times 10^{8}$~\msun\ is located at the centre of the map.  In addition to the target galaxy there are an additional 20 galaxies in the sub-cube.  These galaxies are henceforth referred to as the ``nearby neighbours'' (NN) of the target galaxy; they constitute the first component of the contaminant emission. The flux from these galaxies is represented by the blue spectrum in panel (a) and the total intensity map  in panel (e).  With an associated mass of $5.44\times 10^{9}$~\msun, the nearby neighbours contribute slightly less than 75~per~cent of all the flux in the sub-cube.  The second component of the contaminant emission comes from galaxies that lie outside of the sub-cube, yet  have their flux bleed into it when they have their point-source-like flux distributions convolved with the PSF.  We henceforth refer to these galaxies as the ``distant neighbours'' (DN) of the target galaxy.  The flux from distant neighbours is represented by the red spectrum in panel (a) and the total intensity map  in panel (f).  A contaminant mass of $9.90\times 10^{8}$~\msun\ is contributed by this flux component.  

An interesting point to note from this example is the fact that the mass in the sub-cube associated with the target galaxy, $9.11\times 10^8$~\msun, is only $\sim 65$~per~cent of the evaluated (true) \hi\ mass of the galaxy, $1.51\times 10^{9}$~\msun.  This is expected and is due to the fact that the spatial area over which the sub-cube is extracted is equal on a side to the half-power width of the Gaussian PSF used to spatially smooth the synthetic cube.  A fraction of the target galaxy flux will be distributed by the PSF to distances that place it beyond the spatial extent of the sub-cube.  For the same reason, the total \hi\ mass of $1.26\times 10^{10}$~\msun\  contained in the un-smoothed version of the sub-cube is higher than the mass of $7.34\times 10^9$~\msun\ contained in the smoothed version. 

In the next section we produce co-added spectra by combining many individual spectra extracted from the full-size cube.  Each constituent spectrum is decomposed in the ways presented in this section.  By combining the respective components of the spectra we produce decomposed co-added spectra.  

\begin{figure*}
	\includegraphics[width=2\columnwidth]{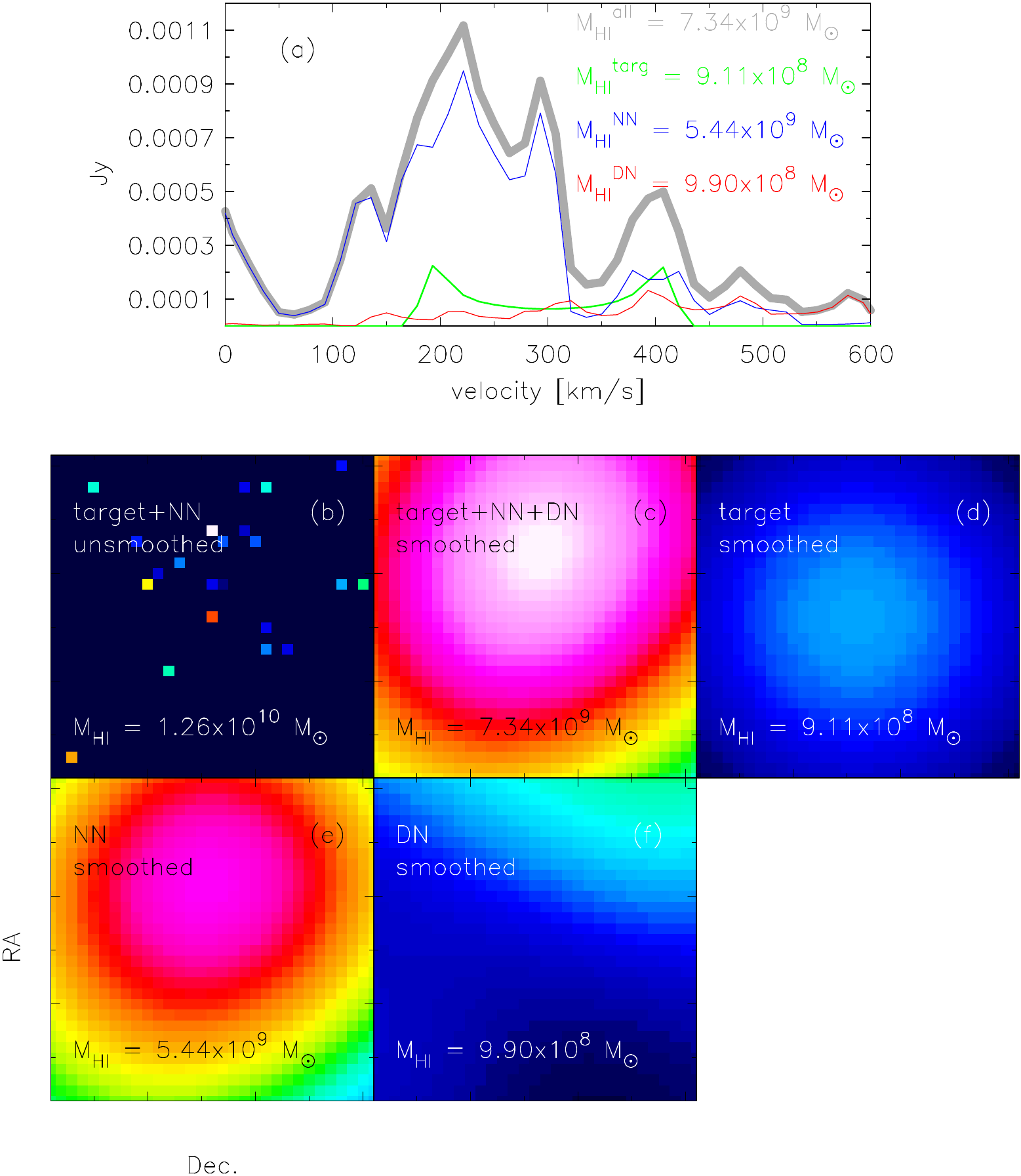}
    \caption{ \hi\ spectra and total intensity maps for the various mass components in a sub-cube extracted from the full-size cube described in Section~\ref{sec:simulations} and presented in Figs.~\ref{fig:lightcone_panel_plot}, \ref{fig:lightcone_channel_maps}, \ref{fig:lightcone_mom0}.  The black crosses in Figs.~\ref{fig:lightcone_channel_maps}, \ref{fig:lightcone_mom0} mark the spatial location of the sub-cube in the full-size cube.  The thick grey curve in panel (a) represents the spectrum consisting of all the flux ($7.34\times 10^9$~\msun) in the sub-cube.  The corresponding total intensity map is shown in panel (c).  The green, blue, and red spectra in panel (a) represent, respectively, the flux in the sub-cube from the target galaxy ($9.11\times 10^8$~\msun), nearby neighbours ($5.44\times 10^{9}$~\msun), and distant neighbours ($9.90\times 10^{8}$~\msun).  The corresponding total intensity maps are shown in panels (d), (e), (f).  The total intensity map shown in panel (b) is that of the sub-cube \emph{before} it is spatially smoothed.  It represents the true brightness distribution of the flux in the sub-cube.  The target galaxy lies at the centre of this map.  Clearly there are 20 nearby neighbour galaxies in the sub-cube.  The colour of each pixel is representative of the total \hi\ flux density of a galaxy, with the total mass for all of the galaxies in the sub-cube being $1.26\times 10^{10}$~\msun.  All of the total intensity maps span 15~arcmin on a side.} 
    \label{fig:15_arcmin_spectra}
\end{figure*}

\section{Mock stacking experiments}
\label{sec:mock_stacks}
\subsection{Low redshift}
In this section we use synthetic cubes to carry out several mock \hi\ stacking experiments.  We consider first a low-redshift stack for which we use the synthetic cube presented and discussed in section~(\ref{sec:simulations}).  The intrinsic (evaluated) properties of the galaxies in this cube are shown in Fig.~\ref{fig:lightcone_panel_plot}.  The redshift range of this cube, $z=0.04 - 0.13$, matches that of the Parkes observations of a 42~deg$^2$ field near the South Galactic Pole from \citet{delhaize_2013}.  

We produce versions of our synthetic cube at spatial resolutions of $\theta=\{2, 3, 4, ..., 13, 14, 15\}$~arcmin.  These resolutions are representative of data from single dish telescopes observing at 21~cm.  Parkes, with its 64~m steerable dish, produces $\sim~$15~arcmin resolution images (e.g.,~\citealt{delhaize_2013}).  Arecibo, with an effective dish diameter of $\sim~200$~m, yields an \hi\ resolution of $\sim~3.5$~arcmin (e.g.,~\citealt{fabello_2011a}).  Future wide-field \hi\ galaxy surveys will have much smaller beam sizes (e.g., $\sim30$~arcsec for the ASKAP \hi\ All-Sky Survey, known as WALLABY), such that source confusion will be less of a concern at low redshifts than what is predicted here by the smallest beam we consider.    Our results are best interpreted in the context of existing single-dish \hi\ galaxy surveys. 

From each of our cubes we extract sub-volumes at the positions of the 3583 galaxies with stellar mass $\mathrm{M}_*\ge 10^{10}$~\msun.  Each sub-volume is spatially delimited by a square-shaped aperture of size $\theta~\mathrm{arcmin}\times \theta~\mathrm{arcmin}$, where $\theta$ is the spatial resolution of the cube from which it is extracted.  All sub-volumes for all cubes are extracted over 13.75~MHz ($\sim~3146$~\kms).  An extracted sub-volume is converted into a spectrum by summing the flux in each spectral channel.

Figure~\ref{fig:spectra_at_various_res}  shows spectra extracted for a particular galaxy from each of our low-redshift synthetic cubes. It is the same galaxy used for the flux confusion example in section~(\ref{sec:flux_decomp}), and its spatial position is marked by the black crosses in Fig.~\ref{fig:lightcone_channel_maps}, \ref{fig:lightcone_mom0}.  It has an intrinsic \hi\ mass of $1.51~\times 10^9$~\msun.  In each panel, the black spectrum represents the flux from all galaxies in the sub-cube, while the green spectrum represents target galaxy flux only.  Furthermore, the spectra are presented as \hi\ mass spectra, generated from the \hi\ flux density spectra using
\begin{equation}
M_{\mathrm{HI}}[\mathrm{M}_{\odot}]=2.36\times 10^5~D_L^2~S_i~dv~(1+z)^{-1},
\end{equation}
where $S_i$ is flux density in units of Jy in channel $i$ of the sub-cube, $dv$ is the velocity width of a channel in \kms, and $D_{\mathrm{L}}$ is the luminosity distance of the target galaxy in Mpc units, and $z$ is its evaluated redshift.

\begin{figure}
	\includegraphics[width=0.9\columnwidth]{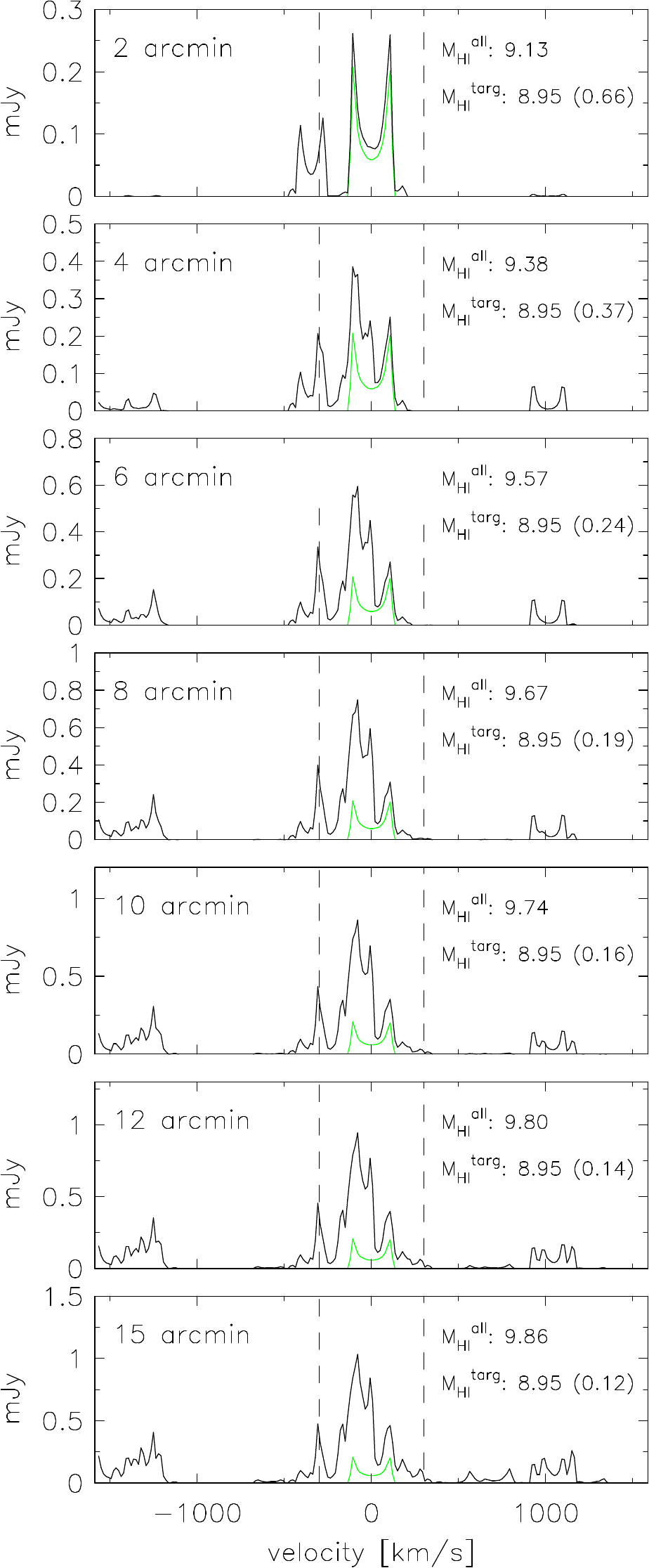}
    \caption{Spectra extracted from the  synthetic cube presented and discussed in section~(\ref{sec:simulations}), ``observed'' at spatial resolutions of 2, 4, 6, 8, 10, 12, 15 arcmin.  The spectra are generated from sub-cubes extracted at the position marked by the cross symbols in Fig.~\ref{fig:lightcone_channel_maps}, \ref{fig:lightcone_mom0}.  Each sub-cube is extracted over a frequency range of 13.75~MHz.  In each panel, the green spectrum represents the flux from the target galaxy only.   The black spectrum in each panel represents all of the flux in the sub-cube (target galaxy plus contaminant flux).  The vertical dashed lines delimit a range of $\pm 300$~\kms\ about the centre of the target spectrum.  $M_{\mathrm{HI}}^{\mathrm{all}}$ is the base 10 logarithm of the \hi\ mass corresponding to all of the flux in the spectrum.  $M_{\mathrm{HI}}^{\mathrm{targ}}$ is the corresponding quantity for the target galaxy flux only.  The number in parentheses is $M_{\mathrm{HI}}^{\mathrm{targ}}$/$M_{\mathrm{HI}}^{\mathrm{all}}$, i.e. the fractional contribution of the target galaxy flux to the total flux in the spectrum.} 
    \label{fig:spectra_at_various_res}
\end{figure}

Clearly evident from the spectra shown in Fig.~\ref{fig:spectra_at_various_res} is the build up of contaminant (non-target) flux with worsening spatial resolution.  This is entirely expected and is due firstly to there being more nearby neighbours in the sub-volumes as they become larger, and secondly to more contaminant emission from distant neighbours bleeding into the sub-volumes as the spatial resolution of the synthetic cubes worsens.  Considering the region $\pm~300$~\kms\ about the centres of the spectra, we see the fractional contribution of target galaxy flux to the total flux drops from 0.66 in the 2~arcmin resolution cube to 0.12 in the 15~arcmin resolution  cube.  It should be noted, however, that the spectra shown in Fig.~\ref{fig:spectra_at_various_res} correspond to a single sight-line, and that large variations can be expected among spectra from different sight-lines.  

The co-added mass spectra  made from the spectra extracted from our low-redshift synthetic cubes are shown in Fig.~\ref{fig:lowz_stacks}.  All of the co-adds have had their total mass (black) decomposed into the contributions from target galaxies (green), nearby neighbours (blue), and distant neighbours (red).  Arguably, the most striking feature of the co-adds is their characteristic shape.  Almost all of them have extended wings beyond $\pm~300$~\kms\  of the centre.  These wings are made up entirely of mass from nearby (blue) and distant (red) neighbours.  All target galaxy mass is contained within $\pm~300$~\kms, which is expected given that we model all galaxies to have a maximum rotation speed less than 300~\kms.  However, within this velocity range there is contaminant mass, too.  Furthermore, for most co-adds it completely overwhelms the target galaxy mass. The target galaxy co-adds (green) have a well-defined double-horn shape.  This is expected for two main reasons: 1) the template rotation curves that we use to construct individual galaxy models are typically flat at outer radii, and 2)  we are considering the ideal case in which there are no uncertainties in the redshifts of the target galaxies.  In reality, optical spectroscopic redshift errors are of order $\sim~30-50$~\kms.  Such uncertainties will lead to an imperfect alignment of target galaxy spectra, resulting in a co-added spectrum that is more boxy in shape over its central $\pm~300$~\kms.  The extent to which the shape of a stacked spectrum is affected by redshift errors is difficult to assess empirically.  Mock stacking experiments based on synthetic data products such as ours can be used to reliably interpret the shape of a co-added spectrum in the context of known redshift uncertainties. 

\begin{figure}
	\includegraphics[width=\columnwidth]{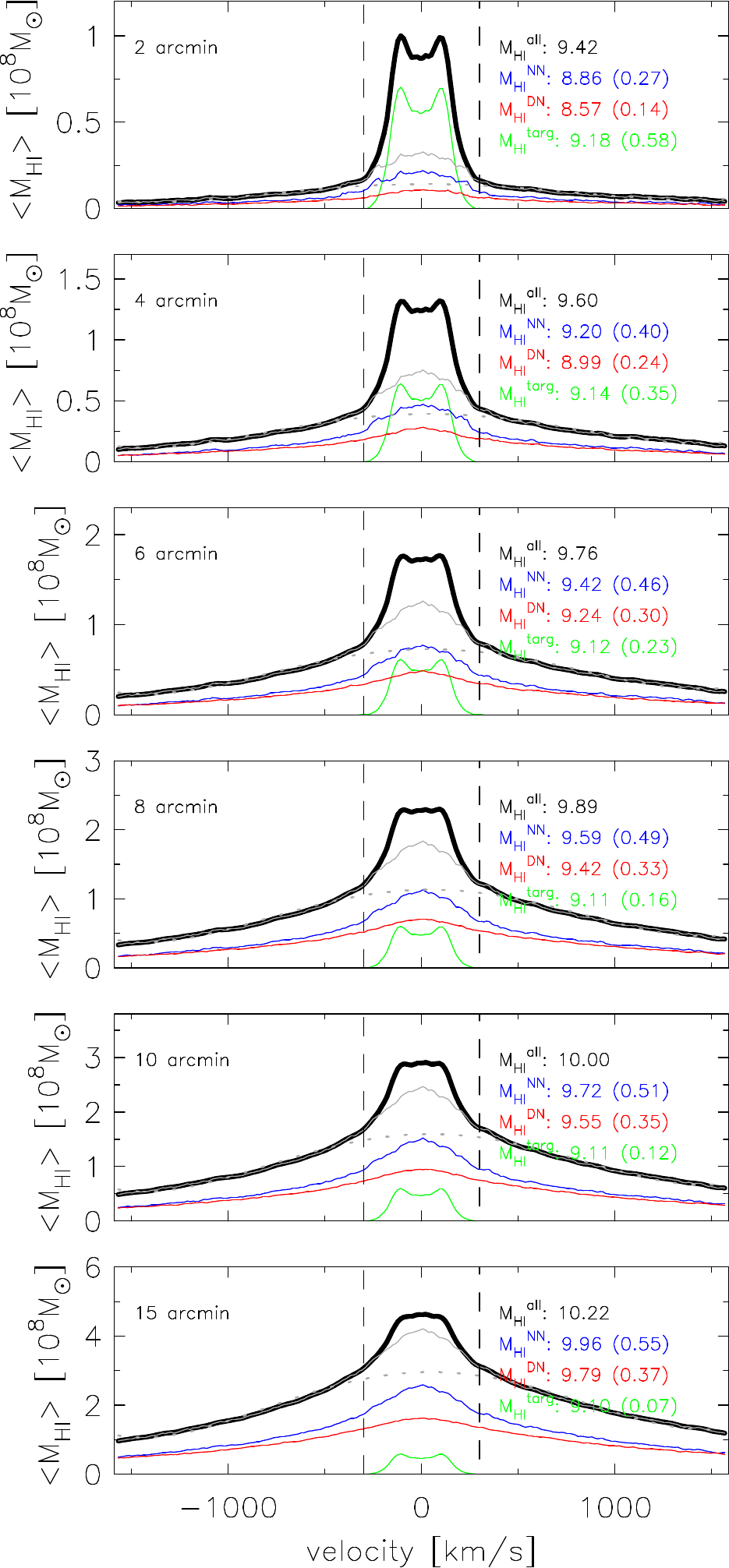}
    \caption{Low-redshift co-added \hi\ mass spectra extracted from synthetic cubes produced at spatial resolutions of 2, 4, 6, 8, 10, 12, 15 arcmin.  3385 \hi\ spectra for galaxies with stellar mass $M_*\ge 10^{10}$~\msun\ contribute to each co-add.  In each panel the solid black curve represents the sum of the target and non-target galaxy mass.  The solid green curve presents the mass from target galaxies only.  The solid blue and red curves represent the mass from nearby and distant neighbours, respectively.  The solid grey curve represents the total non-target mass (the sum of the blue and red curves).  The dotted grey curve represents a fourth-order polynomial fitted to the black curve over the channels outside of the vertical dashed lines that delimit $\pm~300$~\kms\ about the centre of the co-adds.  The first set of numbers in each panel represent the base 10 logarithm of the amount of mass under each curve, integrated between the two vertical dashed lines.  The numbers in parentheses represent the fractional contribution of each mass component to the total co-added mass within $\pm~300$~\kms.}
    \label{fig:lowz_stacks}
\end{figure}

In each panel of Fig.~\ref{fig:lowz_stacks} we specify in parentheses the fractional contribution of each mass component to the total co-added mass.  These quantities are plotted against $\theta$ in Fig.~\ref{fig:mass_fractions_low_z}.  Only the 2~arcmin cube co-add contains more target galaxy mass than non-target galaxy mass.  The fraction of target mass to non-target mass drops quickly with worsening resolution, reaching a value $\lesssim 0.1$ for the 15~arcmin cube co-add.  The blue and red curves in Fig.~(\ref{fig:mass_fractions_low_z}) represent the fractional mass contributions from nearby neighbours (NN) and distant neighbours (DN), respectively.  All co-adds contain more contaminant mass from NN than DN, with the ratio of NN to DN mass always less than 2.  

\begin{figure}
	\includegraphics[width=1\columnwidth]{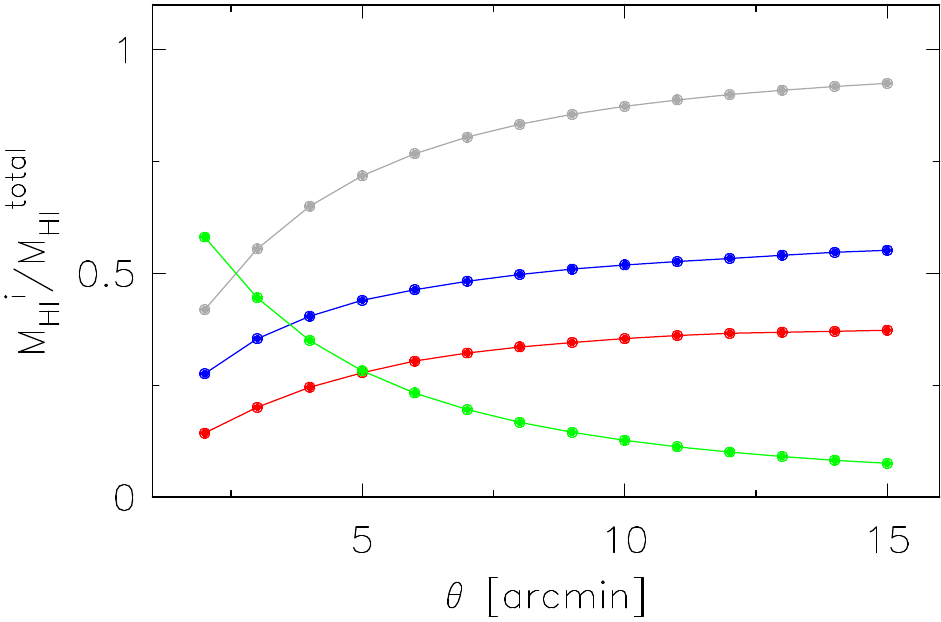}
    \caption{Fractional contributions from target galaxies (green), nearby neighbours (blue) and distant neighbours (red) for the co-added spectra shown in Fig.~(\ref{fig:lowz_stacks}).  The grey curve represent the total mass from nearby and distant neighbours.  All masses correspond to the central $\pm~300$~\kms\ of the co-adds, delimited by the vertical dashed lines in Fig.~\ref{fig:lowz_stacks}.}
    \label{fig:mass_fractions_low_z}
\end{figure}

For each of the co-adds, panels (a)-(c) in Fig.~\ref{fig:analysis_panel_low_z} show as a function of resolution the  average contaminant mass contributed by NN \emph{and} DN galaxies, as well as the separate  contributions from NN and DN galaxies.  A co-added mass spectrum generated from 15~arcmin data can contain $\sim 1.5~\times10^{10}$~\msun\ of contaminant mass per galaxy, on average, if no corrections are applied to the co-add.  Similarly, a co-add based on higher resolution 4~arcmin data can contain more than $10^9$~\msun\ of contaminant mass per galaxy, on average.  These contaminant masses should be compared to the true (evaluated) average \hi\ mass, $2.08\times 10^9$~\msun, of the 3385 target galaxies, shown as a dashed horizontal line in each panel.  In all cases, the average non-target galaxy mass contributing to the spectrum is larger than the average target galaxy mass.  It should be noted that co-added galaxy spectra based on real data will contain more contaminant mass than suggested by the mock co-adds presented in this work.  This is because the \citet{obresch_2014} input catalogue is complete only  for \hi\ masses above $10^8$~\msun.  Most of the \hi\ mass in smaller galaxies is missed by the catalogue.

\begin{figure*}
	\includegraphics[width=2\columnwidth]{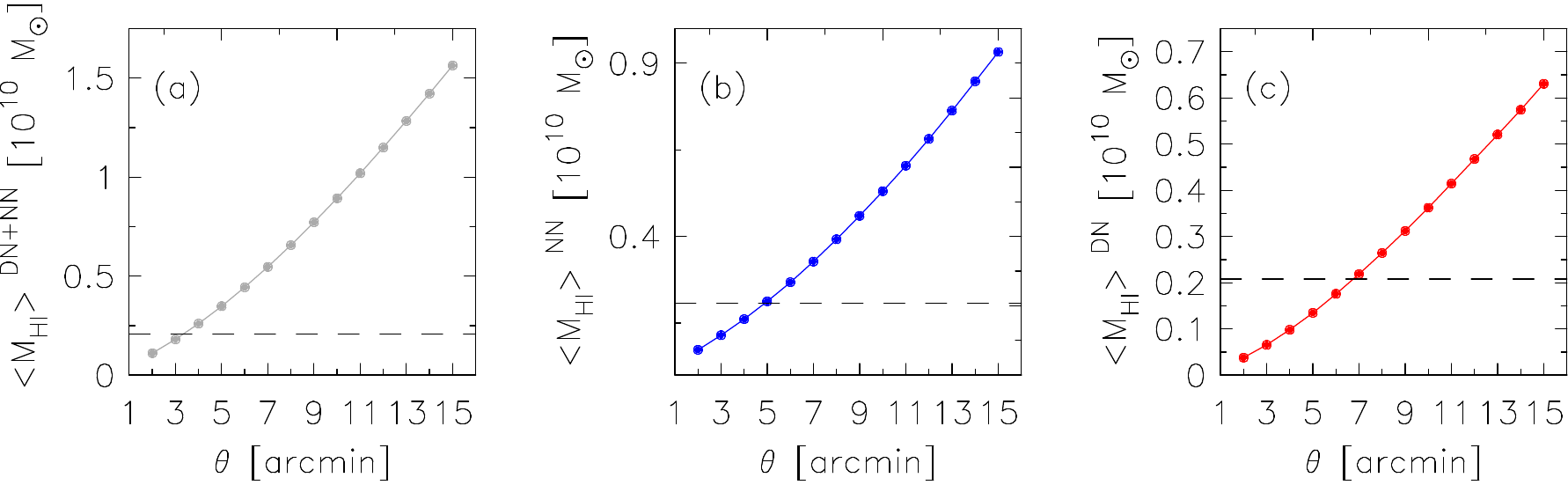}
    \caption{\hi\ masses (averaged per galaxy) for the co-adds shown in Fig.~(\ref{fig:lowz_stacks}), as a function of resolution.  Panel (a): total contaminant mass, panel (b): nearby neighbours contaminant mass, panel (c): distant neighbours contaminant mass.  The black dashed line in each panel represents the true (evaluated) average \hi\ mass, $2.08\times 10^9$~\msun, of the 3385 galaxies that were stacked.}
    \label{fig:analysis_panel_low_z}
\end{figure*}

In an attempt to remove the contaminant flux/mass in the central $\pm~300$~\kms\ of a stacked spectrum, some authors model the contaminant flux in this region by fitting a fourth order polynomial to the extended wings\symbolfootnote[5]{Beyond $\pm~300$~\kms\ of the centre.} of the co-add.  The polynomial is then subtracted from the spectrum in order to remove the contaminant flux.  However, our co-added spectra show this to be an inadequate method.  We have fitted fourth-order polynomials to the wings of the total-mass co-adds (black) shown in Fig.~\ref{fig:lowz_stacks}.  In each panel the fitted polynomial is shown as the dotted grey curve.  An important fact is immediately clear: the polynomial always significantly under-estimates the amount of non-target mass within $\pm~300$~\kms.  This is because the shape of the polynomial over this central velocity range does not match that of the non-target mass.  The polynomial has an almost constant value within $\pm~300$~\kms\ whereas the amount of non-target mass rises sharply towards the centre of the co-add.  A simple polynomial is clearly an inadequate model for non-target mass within $\pm~300$~\kms.  The difference between the solid and dotted grey curves in Fig.~\ref{fig:lowz_stacks} quantifies the amount of non-target mass that survives the polynomial subtraction.  Fig.~\ref{fig:cont} shows the ratio of mass in the polynomial-subtracted total-mass co-add to the mass in the target-galaxy co-add, as a function of resolution.  The ratio is always greater than unity, and increases linearly with worsening resolution.  For a co-added spectrum created from Parkes data ($\sim~15$~arcmin spatial resolution), modelling the contaminant emission as a polynomial yields a new co-add that overestimates the true co-added mass of the target galaxies by a factor $\gtrsim 3.5$. 

\begin{figure}
	\includegraphics[width=\columnwidth]{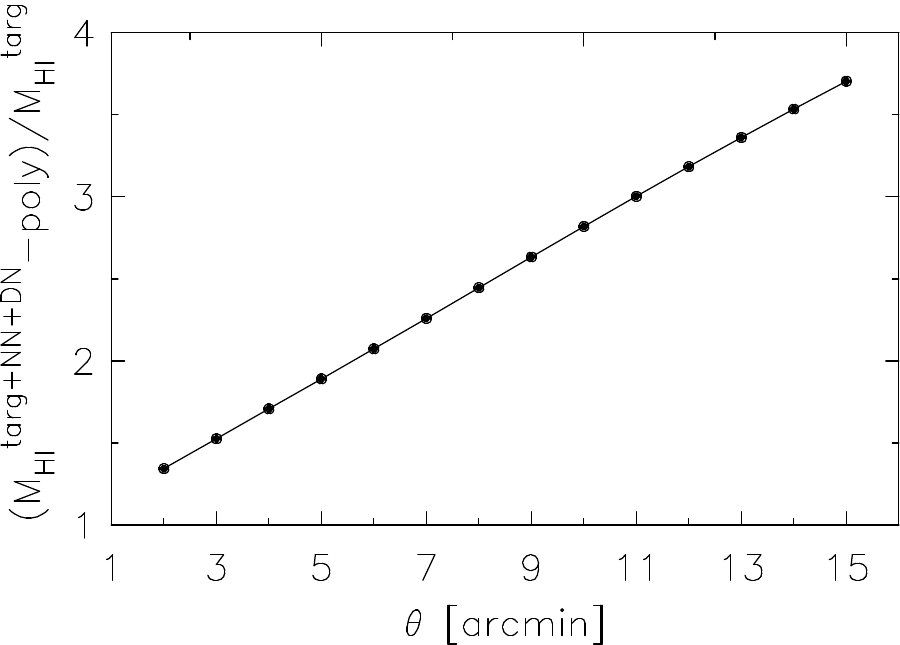}
    \caption{Ratio of total mass to target galaxy mass, as a function of resolution, for the total mass co-adds in Fig.~(\ref{fig:lowz_stacks}) (black curves) that have had their non-target mass within $\pm~300$~\kms\ modelled as a fourth order polynomial (grey dashed curves in Fig.~\ref{fig:lowz_stacks}) and subsequently subtracted.}
    \label{fig:cont}
\end{figure}

\subsection{High redshift}
We now consider a high-redshift stacking experiment based on a cosmological volume spanning a sky area of 0.7~deg$^2$ and the redshift range $z=0.7 - 0.758$.  The corresponding luminosity distance range is $\sim$~4400~-~4850~Mpc, which is very similar in width to that of our low-redshift cube (183~-~633~Mpc).  This volume contains19~922 galaxies with a total \hi\ mass of $1.03\times 10^{13}$~\msun.  Some of the properties of this galaxy sample are shown in Fig.~\ref{fig:highz_lightcone_panel_plot}.

\begin{figure*}
	\includegraphics[width=2\columnwidth]{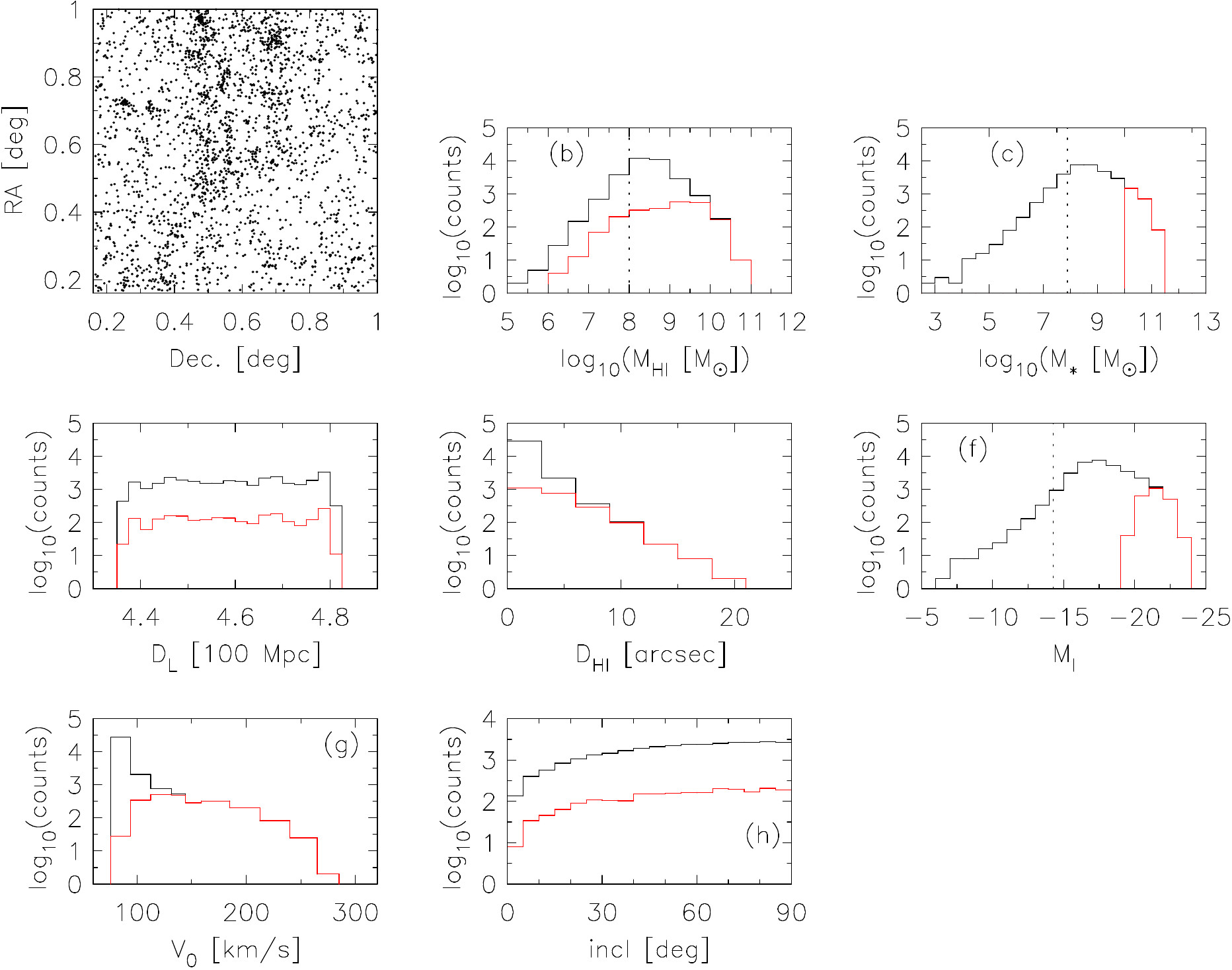}
    \caption{Evaluated galaxy properties for a 0.7 square degree sub-cube of the \citet{obresch_2014} catalogue.  This sub-cube, spanning the redshift range $z=0.7 - 0.758$, contains 19~922 galaxies.  The black histograms correspond to the full sample of 19~922 galaxies, whereas the red histograms are for the subset of 1035 galaxies with stellar mass $M_*>10^{10}$~\msun\ used to produced the co-added spectra shown in Fig.~\ref{fig:highz_stacks}.  Panel~(a): (RA, Dec) positions of galaxies with stellar mass $M_*\ge 10^{9.5}$~\msun.  Panel~(b): distribution of \hi\ masses.  Total \hi\ mass is  $1.03\times 10^{13}$~\msun\ for the full sample.  Panel~(c): distribution of stellar masses.  Panel~(d): distribution of luminosity distances.  Panel~(e): distribution of \hi\ diameters.  Panel~(f): distribution of $I$-band absolute magnitudes.  Panel~(g): distribution of $V_{\mathrm{0}}$ parameters (see eqn~\ref{eq:vrot}), indicative of maximum rotation speed.  Panel~(h): distribution of galaxy inclinations.  The vertical dotted lines in panels b, c, f indicate the masses above which the simulation is complete.}
    \label{fig:highz_lightcone_panel_plot}
\end{figure*}

We use the evaluated galaxy properties to again produce a suite of simulated cubes, this time having spatial resolutions of $\theta~=~\{18.8, 37.7,56.6,75.5,94.3,141.5\}$~arcsec.  For the redshift range $z=0.7 - 0.758$, these spatial resolutions yield a similar number of galaxies per beam as the resolutions of 2, 4, 6, 8, 10, 15 arcmin for the low-redshift cubes.  These resolutions are representative of \hi\ imaging sets from current and future high-redshift galaxy surveys.  For example, the results from a pilot for an \hi\ deep survey of the COSMOS field done with the VLA in $B$-configuration yielded \hi\ maps with spatial resolution $\sim 5$~arcsec \citep{chiles_pilot}.  The LADUMA survey \citep{LADUMA}, to be carried out on the 64-element MeerKAT array, will yield \hi\ image sets of resolution $\sim 10 - 20$~arcsec, whereas data from the Widefield ASKAP L-band Legacy All-sky Blind surveY (WALLABY) will be at a resolution of $\sim~30$~arcsec.

All of our synthetic cubes have a channel width of 62.5~kHz ($\sim~22.8$~\kms\ at the mean redshift of 0.728) and a pixel scale of 3~arcsec.  For each simulated cube we extract sub-volumes at the positions of the 1035 galaxies with stellar mass $M_*\ge 10^{10}$~\msun.  Each sub-volume is spatially delimited by a square-shaped aperture of size $\theta~\mathrm{arcsec}\times \theta~\mathrm{arcsec}$, where $\theta$ is the spatial resolution of the cube from which it is extracted.  All sub-volumes for all cubes are extracted over 8.125~MHz ($\sim~2969$~\kms).  Each sub-volume is converted into a spectrum by summing the flux in each spectral channel.  

Figure~\ref{fig:highz_stacks} shows the co-added mass spectra made from the spectra extracted from our high-redshift synthetic cubes.  All co-adds have again had their total mass spectra (black) decomposed into contributions from target galaxies, nearby neighbours and distant neighbours (green, blue and red spectra, respectively).  The general trends for the high-redshift co-adds are very similar to those of the low-redshift co-adds: the target galaxy mass fractions steadily decrease with worsening spatial resolution, while the contributions from non-target galaxies increase.  However, at 18.8~arcsec resolution, target galaxy mass constitutes slightly less than 70~per~cent of the total co-added mass within $\pm~300$~\kms\ of the spectral centre.  This should be compared to the corresponding mass fraction for the 2~arcmin low-redshift cube - its target galaxies constitute only 58~per~cent of the total co-added mass.  This result is, in fact, true for the co-adds from all of our high-redshift cubes - each contains a larger fraction of target galaxy mass than its low-redshift counterpart.  This is clearly shown in Fig.~\ref{fig:mass_fractions_high_z}, which plots as filled circles the fractional contributions of the various mass components for the high-redshift cubes, and as dotted curves the  fractional contributions for the corresponding low-redshift cubes.  While the two sets of co-added spectra contain similar fractional mass contributions from distant neighbours, the low-redshift stacks have significantly higher contributions from nearby neighbours. Hence, overall they have larger non-target galaxy mass fractions.  In terms of source confusion rates, these results show that interferometric \hi\ spectral line stacking experiments carried out at redshifts of $z\sim 0.7$ will yield more reliable results than low-redshift experiments carried out with single dish telescopes.

\begin{figure}
	\includegraphics[width=1\columnwidth]{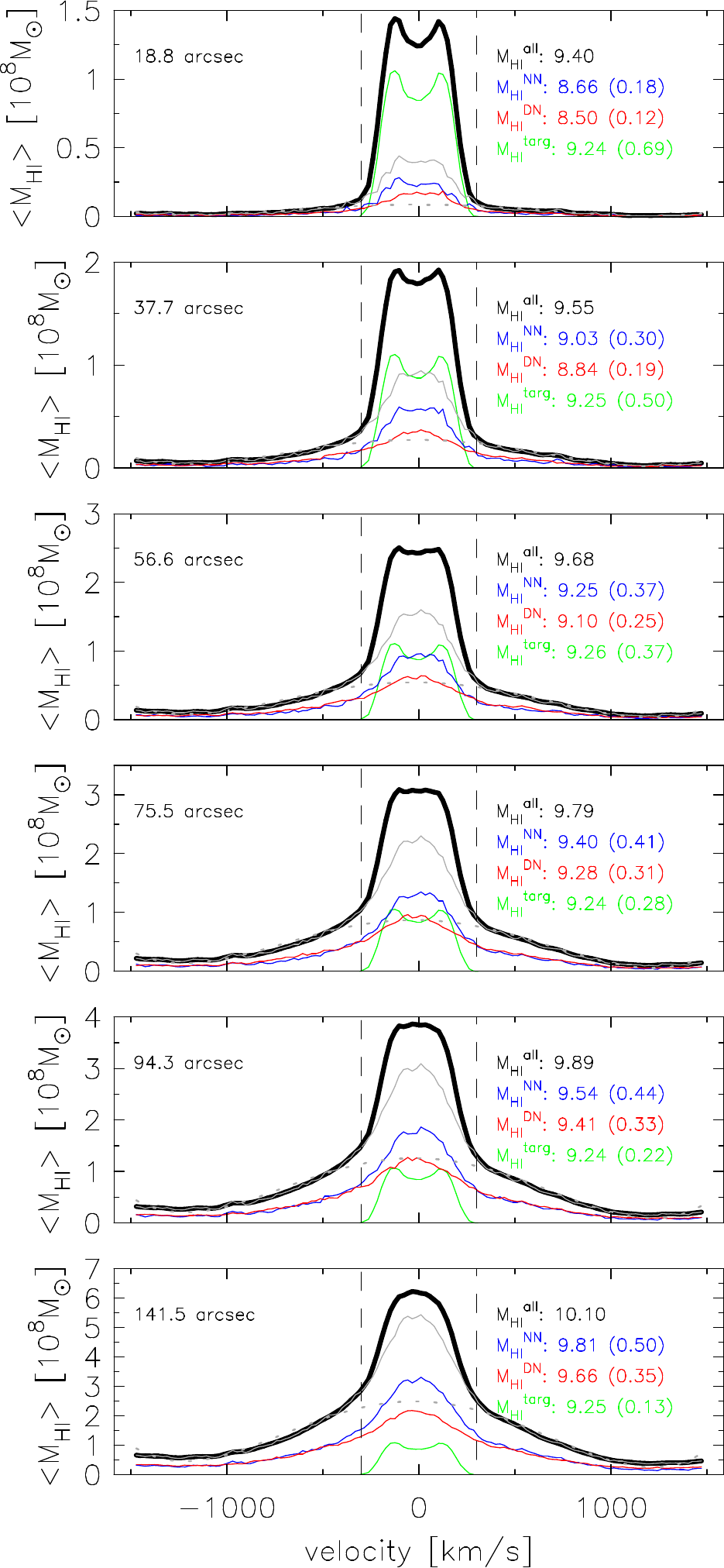}
    \caption{High-redshift co-added \hi\ mass spectra extracted from synthetic cubes produced at spatial resolutions of 18.8, 37.7, 56.6, 75.5,  94.3,  141.5~arcsec.  1035 \hi\ spectra for galaxies with stellar mass $M_*\ge 10^{10}$~\msun\ contribute to each co-add.  See Fig.~(\ref{fig:lowz_stacks}) caption for further details.}
    \label{fig:highz_stacks}
\end{figure}

\begin{figure}
	\includegraphics[width=1\columnwidth]{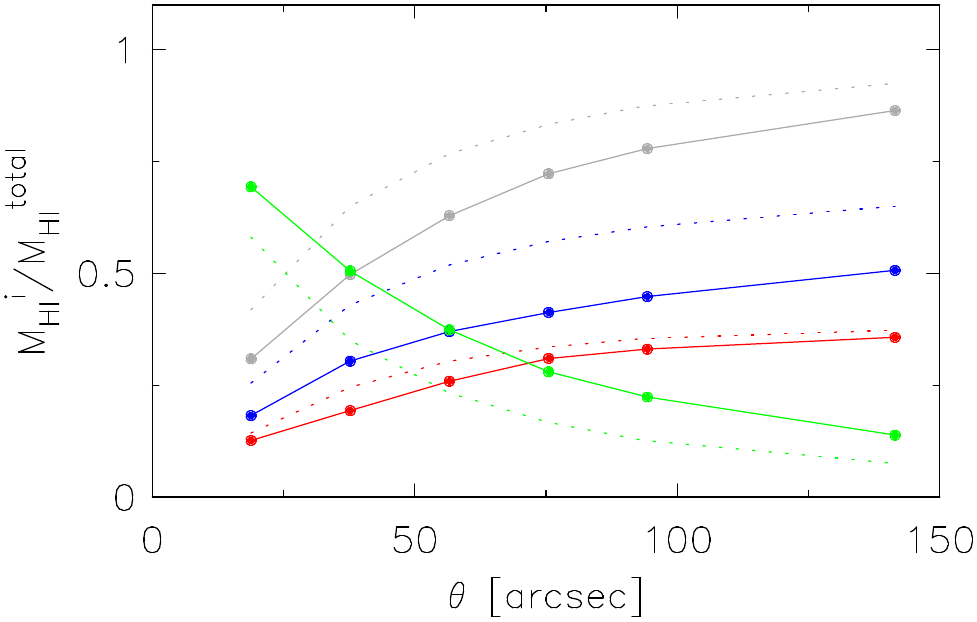}
    \caption{Shown as filled circles are the fractional contributions from target galaxies (green), nearby neighbours (blue) and distant neighbours (red) for the co-added spectra shown in Fig.~(\ref{fig:highz_stacks}).  The grey curve represent the sum of the red and blue curves.  All masses correspond to the central $\pm~300$~\kms\ of the co-adds, delimited by the vertical dashed lines.  Represented by the dotted curves are the fractional mass contributions for the low-redshift stacks at the corresponding resolutions of 2, 4, 6, 8, 10~arcmin.}
    \label{fig:mass_fractions_high_z}
\end{figure}

For each of the co-adds, panels (a) - (c) in Fig.~\ref{fig:analysis_panel_high_z} show as a function of resolution the average contaminant mass contributed by NN \emph{and} DN, as well as the separate contributions from NN and DN galaxies.  The total contaminant mass per galaxy varies from $\sim~0.1 - 1.1\times 10^{10}$~\msun\ for  spatial resolutions from 18.8 to 141.5~arcsec.  Also shown in the panels, as dotted curves, are the corresponding masses for the low-redshift cubes.  The high-redshift stacks all have a lower average contaminant mass than their low-redshift counterparts.  

\begin{figure*}
	\includegraphics[width=2\columnwidth]{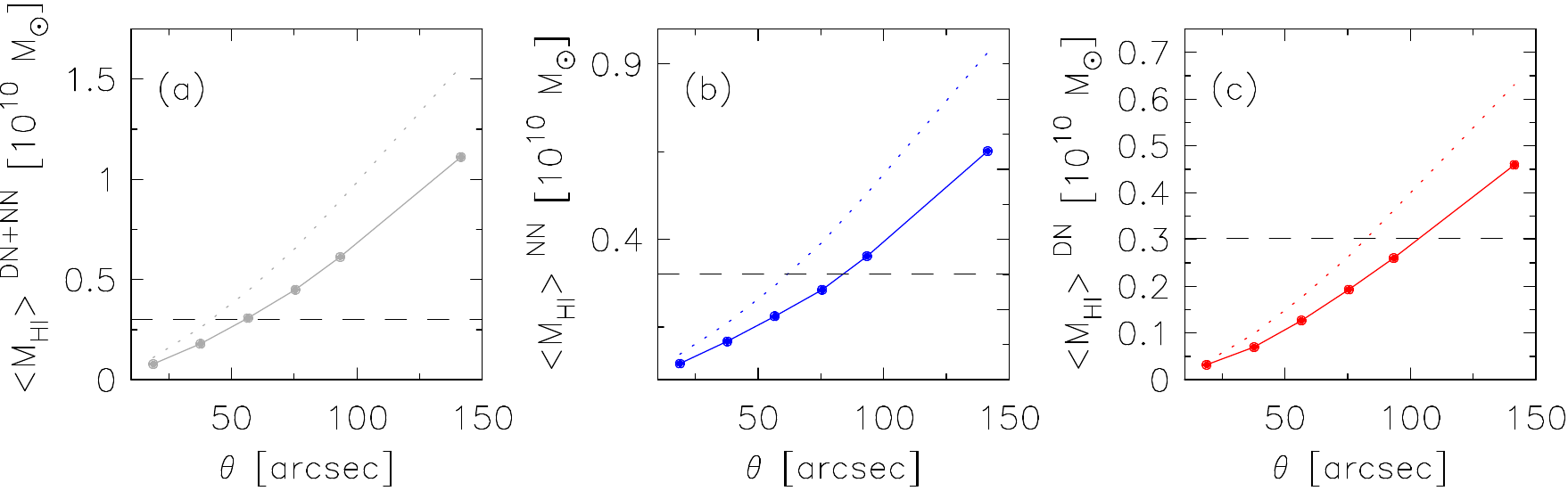}
    \caption{\hi\ masses (averaged per galaxy) for the co-adds shown in Fig.~(\ref{fig:highz_stacks}), as a function of resolution.  Panel (a): total contaminant mass, panel (b): nearby neighbours contaminant mass, panel (c): distant neighbours contaminant mass.  The dotted lines represent masses from the corresponding low-redshift co-adds.  The black dashed line in each panel represents the true (evaluated) average \hi\ mass, $3.02\times 10^9$~\msun, of the 1035 galaxies that were stacked.}
    \label{fig:analysis_panel_high_z}
\end{figure*}

\section{Discussion}
\label{sec:discussion}
\subsection{Theoretical vs observational comparisons}
The mock stacking experiments presented in this work show contaminant emission from non-target galaxies  significantly contributes to the co-added spectra of both low and high-redshift galaxy samples.  While our approach to quantifying the amount of contaminant flux in a co-added spectrum is highly theoretical in nature, other authors have done so from a purely observational perspective.  

\citet{Jones_2015b} use an analytic model to predict the amount of confused mass in a co-added spectrum for a generic \hi\ survey.  Their model uses the ALFALFA correlation function \citep{Papastergis_2013} and and the ALFALFA measurement of the \hi\ mass function in the local Universe \citep{Martin_2010}.  They use their model to estimate an average contaminant mass per galaxy of $1.3\times 10^{10}$~\msun\  for the South Galactic Pole (SGP) stacked \hi\ profile of \citet{delhaize_2013} who used Parkes data for  3277 galaxies in the redshift range $z=0.04 - 0.13$.  The \citet{Jones_2015b} estimate is based on a beam size of 15.5~arcmin and considers only the cylindrical volume spanning a velocity range of 600~\kms.  Their estimate can therefore be directly compared to our determination for the average contaminant mass in our co-added spectrum based on the spectra extracted from our 15~arcmin low-redshift cube (Fig.~\ref{fig:lowz_stacks}, bottom left panel).  Our simulation suggests a value of $1.5\times 10^{10}$~\msun, of which $\sim~60$ and 40~per~cent is from nearby and distant neighbours, respectively.  The results from our purely theoretical approach therefore compare very favourably to those based on observational data, at least for the local Universe.  

\citet{Jones_2015b} also make predictions for the amount of contaminant mass, on average, in a co-added spectrum made from a generic survey in the redshift range $z=0-1.4$.   They do this by assuming $\Omega_{\mathrm{HI}}$ to be fixed to the $z=0$ value of $4.3\times 10^{-3}$ from \citep{Martin_2010}, which is based on the 40~per~cent ALFALFA survey, and they again use the ALFALFA 2D correlation function.  They estimate an average confused mass of $\sim10^9$~\msun\ at $z\sim 0.7$ for a 20~arcsec beam width.  This result can be directly compared to our estimate of the total contaminant mass in our co-added spectrum based on the spectra from our high-redshift 18.8~arcsec resolution cube.  Our simulation yields an average contaminant mass of $7.8\times 10^8$\msun, in good agreement with the observation-based result from \citet{Jones_2015b}.  

Our simulations can also be used to gain additional insights into the more detailed characteristics of the contaminant emission.  The \citet{obresch_2014} simulations on which our synthetic cubes are based use a set of physical prescriptions to model the redshift evolution of \hi\ and CO in galaxies.  These physically motivated prescriptions allow for the successful reproduction of several measured galactic properties at $z=0$.  Our simulated data products incorporate the evolutionary effects from the \citet{obresch_2014} simulations; we are not restricted to extrapolating $z=0$ measurements of galaxies to higher redshifts.  In future work, we aim to study the effects of cosmic variance on the results of high-redshift stacking experiments, the sorts of which will be routinely carried out by LADUMA.

\subsection{Implications for cosmic gas density overestimates}
The results from our mock stacking experiments hold significant implications for the results of previous and future \hi\ stacking experiments.  The take-home message is that most co-added spectra suffer markedly from source confusion, and are most times completely dominated by contaminant flux.  We have already shown a typical Parkes co-add to contain only $\sim~7$~per~cent target-galaxy emission if no corrections are applied.  Modelling the contaminant flux in the co-add as a polynomial still leaves $\sim~3.5$ times  more confused mass than target-galaxy mass.  

A corresponding  correction factor can be generated for a typical stacking experiment based on ALFALFA data.  We generated an ALFALFA-like synthetic cube with a spatial resolution of $\sim~3.3\times~3.8$~arcmin$^2$ spanning the redshift range $z=0.025$~-~0.05, and stacked the  spectra of galaxies with stellar mass $M_*\ge 10^{10}$~\msun.  Despite having a  much better spatial resolution than Parkes \hi\ data, as well as probing a closer redshift range, our co-added spectrum based on 124 galaxy spectra contains only $\sim~67$~per~cent target galaxy mass. 

Several  \hi\ spectral line stacking experiments have been carried out in recent years.  \citet{chengular_2001}  used the Australia Telescope Compact Array to observe Abell~3128 at $z=0.06$, \citet{lah_2007}  used observations from the Giant Metrewave Radio Telescope (GMRT) to measure the \hi\ content of star-forming galaxies at $z=0.24$, \citet{lah_2009}  used GMRT data to measure the \hi\ content of 324 galaxies around Abell~370 at $z=0.37$, \citet{fabello_2011a} stacked the ALFALFA \hi\ spectra of galaxies in the range $z=0.025-0.05$, \citet{delhaize_2013} stacked the Parkes spectra of galaxies with redshifts $z<0.13$, and \citet{gereb_2013}  used data taken with the Westerbork Synthesis Radio Telescope (WSRT) to study the relation between \hi\ and infrared/optical properties of galaxies out to $z=0.09$.   The combined results of these experiments are generally regarded as providing evidence of no evolution of $\Omega_{\mathrm{HI}}$ out to $z\sim 2$.  Our synthetic data products can be used to accurately quantify any potential over-estimates of average galaxy mass from each of these experiments.  If this collection of $\Omega_{\mathrm{HI}}$ measurements is properly corrected for the effects of source confusion, they may indeed provide us with some evidence of an evolving cosmic gas density.  

\section{Conclusions}
\label{sec:conclusions}

In this work we have presented our methods of producing synthetic \hi\ data cubes based on the galaxy catalogues of \citet{obresch_2014}.  Our methods use the evaluated properties to produce realistic models of the spatial and spectral distribution of \hi\ in galaxies spanning large cosmological volumes.  Each galaxy is  modelled using unique parameterisations for its rotation curve and \hi\ mass distribution.  Our synthetic cubes are spatially smoothed with Gaussian convolution kernels in order to mimic the effects of radio telescope point spread functions.  

A very powerful application of our simulated cubes is to calculate, with a high degree of accuracy and reliability, the effects of source confusion in co-added \hi\ spectra.  We have used several noise-free synthetic \hi\ data cubes to carry out mock \hi\ stacking experiments.  For each galaxy in the cube with stellar mass M$_*\ge 10^{10}$~\msun\ we extract a small sub-volume.  From each sub-volume we generate an \hi\ spectrum.  When we spectrally align and then co-add our \hi\ spectra we obtain stacked spectra that have large wings extending to velocities well beyond $\pm~300$~\kms\ of the centre of the co-add.  These wings are due to the source confusion.  A unique advantage offered by our synthetic data products is the ability to decompose the co-added spectra into contributions from the target galaxies of interest, as well as contaminant contributions from other nearby non-target galaxies. In all cases we find the contaminant emission to constitute much of the co-added \hi\ mass within $\pm~300$~\kms.  At low redshifts, for spatial resolutions ranging from 2 to 15 arcmin, the fractional contribution of target galaxy mass to a co-add decreases from $\sim~0.58$ to as little as $\sim~0.07$, showing that the total mass of a stacked spectrum is a significant over-estimate of the true total mass of the galaxy sample of interest.  For a Parkes-like stacking experiment in the redshift range $z=0.04$~-~0.13 we calculate an average contaminant mass per galaxy of $1.5\times10^{10}$~\msun, very similar to an estimate of  $1.3\times10^{10}$~\msun\ from \citet{Jones_2015b} who use an empirical method based on the ALFALFA correlation function.   

Because the \citet{obresch_2014} simulations predict the cosmic evolution of the gas in galaxies, our products automatically incorporate such effects, too.  They can therefore be used to reliably interpret the results of high-redshift \hi\ stacking experiments.  We have also carried out high-redshift ($z~\sim0.7$) \hi\ stacking experiments using our synthetic \hi\ cubes.  The results suggest that interferometric \hi\ stacking experiments such as LADUMA on the MeerKAT array with a spatial resolution of 10~-~30~arcsec will not be dominated by source confusion at a redshift $z\sim0.7$. Rather, they will have have $\sim~69$~per~cent of the flux in their co-added spectra constituted by target galaxy emission.  However, LADUMA plans to stack the spectra of galaxies well beyond $z=1$, which may lead to significantly higher confusion rates.   The precise correction factors that will need to be applied to the results of such stacking experiments in order to allow for reliable measures of $\Omega_{\mathrm{HI}}$ at high redshifts can be calculated using our simulations.

Given the very large confusion rates we have calculated for single dish \hi\ stacking experiments for the nearby Universe, we suggest the possibility that $\Omega_{\mathrm{HI}}$ measurements from previous \hi\ stacking experiments are likely overestimates of the true value/s.  Using our simulations to properly calculate the extents to which existing $\Omega_{\mathrm{HI}}$ measurements overestimate the true value/s could lead to a view in which there is indeed evidence of redshift evolution in $\Omega_{\mathrm{HI}}$ for $z\lesssim 1$.

\section{Acknowledgements}
\label{sec:Acknowledgements}
ECE thanks the South Africa SKA project for supporting this research.  All authors acknowledge funding received from the South African National Research Foundation.  AJB acknowledges support from a 2015-2016 Fulbright Scholarship.  Special thanks are extended to Danail Obreschkow for providing us with galaxy catalogues from \citet{obresch_2014}.  All authors thank the anonymous referee for providing very useful feedback that improved the quality of the paper.

\bibliographystyle{mnras}



\bsp	
\label{lastpage}
\end{document}